\documentclass[aps,prc,10pt,twocolumn,superscriptaddress,showpacs]{revtex4-1}

\usepackage{amsfonts,amsmath,amssymb,bm,color,graphicx,algorithm,algorithmic}

\newcommand{\eq}[1]{(\ref{#1})}
\newcommand{\mtx}[2]{\left[\begin{array}{#1} #2 \end{array}\right]}

\newcommand{\lw}[1]{\smash{\lower2.ex\hbox{#1}}}

\newcommand{\lA}{\leftarrow} 
\def\Mmaru#1{{\ooalign{\hfil#1\/\hfil\crcr\raise.167ex\hbox{\mathhexbox 20D}}}}

\newcommand{\RR}{\mathbb{R}}

\newcommand{\cN}{{\cal N}}

\newcommand{\cP}{{\cal P}}

\newcommand{\veps}{\varepsilon}

\begin{document}

\title{A machine learning-based selective sampling procedure \\
for identifying the low energy region in a potential energy surface: \\
a case study on proton conduction in oxides}
\author{Kazuaki \surname{Toyoura}}
\thanks{Equally contributed as the second author; Corresponding author}
\email{toyoura@numse.nagoya-u.ac.jp}
\affiliation{Department of Materials Science and Engineering, Nagoya University, Nagoya 464-8603, Japan}
\author{Daisuke \surname{Hirano}}
\thanks{Equally contributed as the first author}
\affiliation{Department of Engineering, Nagoya Institute of Technology, Nagoya 466-8555, Japan}
\author{Atsuto \surname{Seko}}
\affiliation{Department of Materials Science and Engineering, Kyoto University, Kyoto 606-8501, Japan}
\affiliation{Center for Elements Strategy Initiative for Structure Materials (ESISM), Kyoto University, Kyoto 606-8501, Japan}
\affiliation{Center for Materials Research by Information Integration, National Institute for Materials Science, Tsukuba 305-0047, Japan}
\author{Motoki \surname{Shiga}}
\affiliation{Department of Electrical, Electronic and Computer Engineering, Gifu University, Gifu, 501-1193, Japan}
\author{Akihide \surname{Kuwabara}}
\affiliation{Nanostructures Research Laboratory, Japan Fine Ceramics Center, Nagoya, 456-8587, Japan}
\author{Masayuki \surname{Karasuyama}}
\affiliation{Department of Engineering, Nagoya Institute of Technology, Nagoya 466-8555, Japan}
\affiliation{Center for Materials Research by Information Integration, National Institute for Materials Science, Tsukuba 305-0047, Japan}
\author{Kazuki \surname{Shitara}}
\affiliation{Center for Elements Strategy Initiative for Structure Materials (ESISM), Kyoto University, Kyoto 606-8501, Japan}
\affiliation{Nanostructures Research Laboratory, Japan Fine Ceramics Center, Nagoya, 456-8587, Japan}
\author{Ichiro \surname{Takeuchi}}
\thanks{Corresponding author}
\email{takeuchi.ichiro@nitech.ac.jp}
\affiliation{Department of Engineering, Nagoya Institute of Technology, Nagoya 466-8555, Japan}
\affiliation{Center for Materials Research by Information Integration, National Institute for Materials Science, Tsukuba 305-0047, Japan}

\date{\today}


\begin{abstract}
In this paper, we propose a selective sampling procedure to preferentially evaluate a potential energy surface (PES) in a part of the configuration space governing a physical property of interest. The proposed sampling procedure is based on a machine learning method called the \emph{Gaussian process (GP)}, which is used to construct a statistical model of the PES for identifying the region of interest in the configuration space. We demonstrate the efficacy of the proposed procedure for atomic diffusion and ionic conduction, specifically the proton conduction in a well-studied proton-conducting oxide, barium zirconate $({\rm BaZrO_3})$. The results of the demonstration study indicate that our procedure can efficiently identify the low-energy region characterizing the proton conduction in the host crystal lattice, and that the descriptors used for the statistical PES model have a great influence on the performance.
\end{abstract}

\maketitle

\section{Introduction}
\label{sec:introduction}

The concept of a potential energy surface (PES) is of great importance for a fundamental understanding of physical phenomena and properties, such as molecular and lattice vibrations, thermodynamic properties, chemical reactions, atomic diffusion and ionic conduction \cite{lombardo1991review,starr2001thermodynamic,fernandez2006modeling,huang2008new,hellman2011lattice,thomas2013finite}.
A PES is generally described as a function of $3n$ degrees of freedom in an $n$-atom system, but it is often simplified by focusing only on reduced degrees of freedom that govern the physical property of interest \cite{reuter2006first,corral2006influence,rosini2008adsorption}.
In this simplification, the rest of the degrees of freedom are effectively incorporated into the reduced configuration space as the minimized energy or statistical average.
For example, when tracing the movement of an diffusing atom in a solid, the host lattice structure is optimized with respect to the energy, so that the movements of the other atoms forming the host crystal lattice are implicitly treated \cite{corral2006influence,rosini2008adsorption}.
This approach is useful for evaluating physical properties mainly governed by the motions of a few atoms, such as atomic diffusion, ionic conduction, and elementary steps of chemical reactions.

In the simplified approach, the entire PES is evaluated for the reduced degrees of freedom, generally by density functional theory (DFT) calculations.
A fine grid is usually introduced in the reduced configuration space to evaluate the potential energies (PEs) at uniformly-distributed points in the configuration space.
The most accurate method is for the PEs at all the grid points to be fully computed with the host lattice structure optimization, which requires that a great amount of computational resources are devoted.
However, one should note that the entire PES is not necessarily required for understanding physical phenomena and properties. 
In the case of atomic diffusion and ionic conduction, a diffusion or a conduction process is fully characterized by a connected low-energy migration pathway throughout the host lattice, meaning that we need not to explore grid points with high PEs irrelevant to the migration pathway \cite{vineyard1957frequency,henkelman2000climbing}. 
Therefore, it is sufficient to know only a region exhibiting low energy, called the \emph{low PE region}, for finding the migration pathway.

In this paper, we propose a selective sampling procedure on the basis of a machine learning technique for evaluating all points in the low PE region containing the migration pathway without evaluating the entire PES. Here we introduce a statistical PES model to sample points in the low PE region efficiently and to ensure that all points in the low PE region have already been sampled without evaluating the entire PES. In particular, the latter is difficult to achieve without using a statistical model. This procedure is generally applicable to any case of finding points in a region of interest, if the region can be mathematically defined. For demonstrating the performance of our selective sampling procedure, we apply it to proton conduction in a well-studied proton-conducting oxide, barium zirconate $({\rm BaZrO_3})$ \cite{bjorketun2005kinetic,iwahara1993protonic,munch2000proton,islam2000ionic}. Specifically, a fine grid is introduced in the crystal lattice of the host oxide, and several sampling procedures are compared in terms of efficiency for finding all the grid points with PEs less than a given PE threshold. 

Note that the proposed procedure is based on a different concept than those of previously reported machine-learning-based material science studies \cite{gu2006using,behler2007generalized,xu2009two,bartok2010gaussian,setyawan2011high,rupp2012fast,saad2012data,fujimura2013accelerated,montavon2013machine,seko2014machine,seko2014sparse,schutt2014represent,dey2014informatics,ghiringhelli2015big,faber2015crystal,meredig2014combinatorial,faber2015machine,Ramakrishnan15}. Specifically, the goal of our procedure is NOT estimating the entire PES from a sparse sampling using regression methods \cite{behler2007generalized,bartok2010gaussian,seko2014sparse}. In the present study, machine-learning techniques are used to identify all points related to the partial PES governing a physical property of interest. Consequently, the present procedure identifies the complete partial PES. Although a hybrid method of the proposed and previous procedures should be more efficient than the present approach for evaluating a reasonable partial PES, this paper focuses on the procedure for finding all points related to the partial PES for simplicity.

\section{Gaussian process Model-based Selective Sampling Procedure}
\label{sec:method}

In this section, we present the proposed selective sampling procedure in detail. Although the procedure is fairly general and can be used in a variety of problems, we describe it here in the context of low PE region detection problems. Given a fine set of grid points in the configuration space, the goal is to efficiently identify the subset of the grid points at which the PEs are relatively low. The procedure has the following three key features. (1) First, a statistical model of the PES is developed as a \emph{Gaussian process (GP)} \cite{williams2006gaussian,stein2012interpolation}. The model is iteratively updated by repeating (i) selectively sampling a point at which the PE is predicted to be low, and (ii) updating the model based on the newly calculated PE value at the sampled point. (2) Second, the statistical PES model is used for identifying the subset of grid points at which the PEs are relatively low. We introduce a novel selection criterion for this purpose, because the single global minimum or maximum point (not a subset) has been targeted in GP applications so far. (3) Finally, the procedure allows us to estimate how many low PE points remain unsampled, i.e., lets us know when the sampling should be stopped.

\begin{figure*}[p]
 \begin{center}
  \begin{tabular}{ccc}
   \multicolumn{3}{c}{\includegraphics[width=0.75\textwidth]{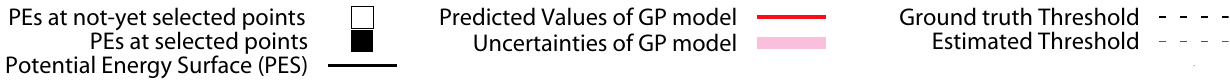}} \\
   \includegraphics[width=0.33\textwidth]{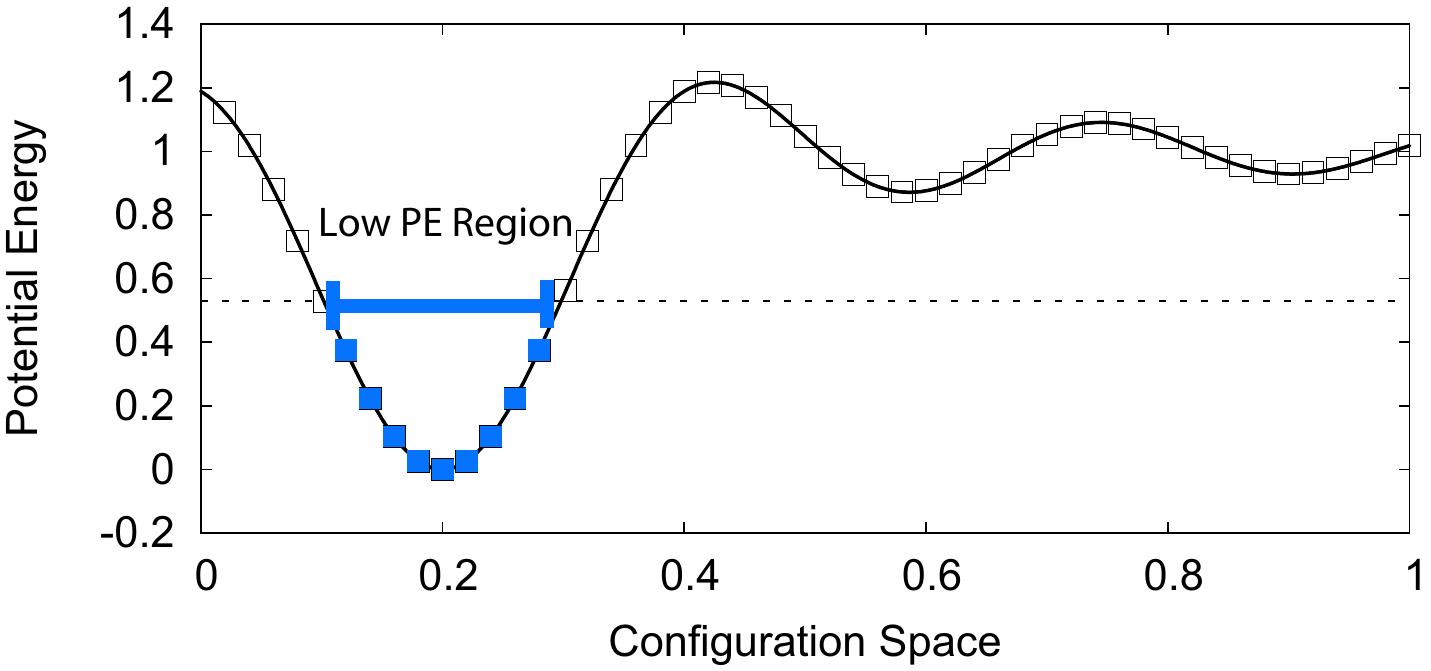} &
   \includegraphics[width=0.33\textwidth]{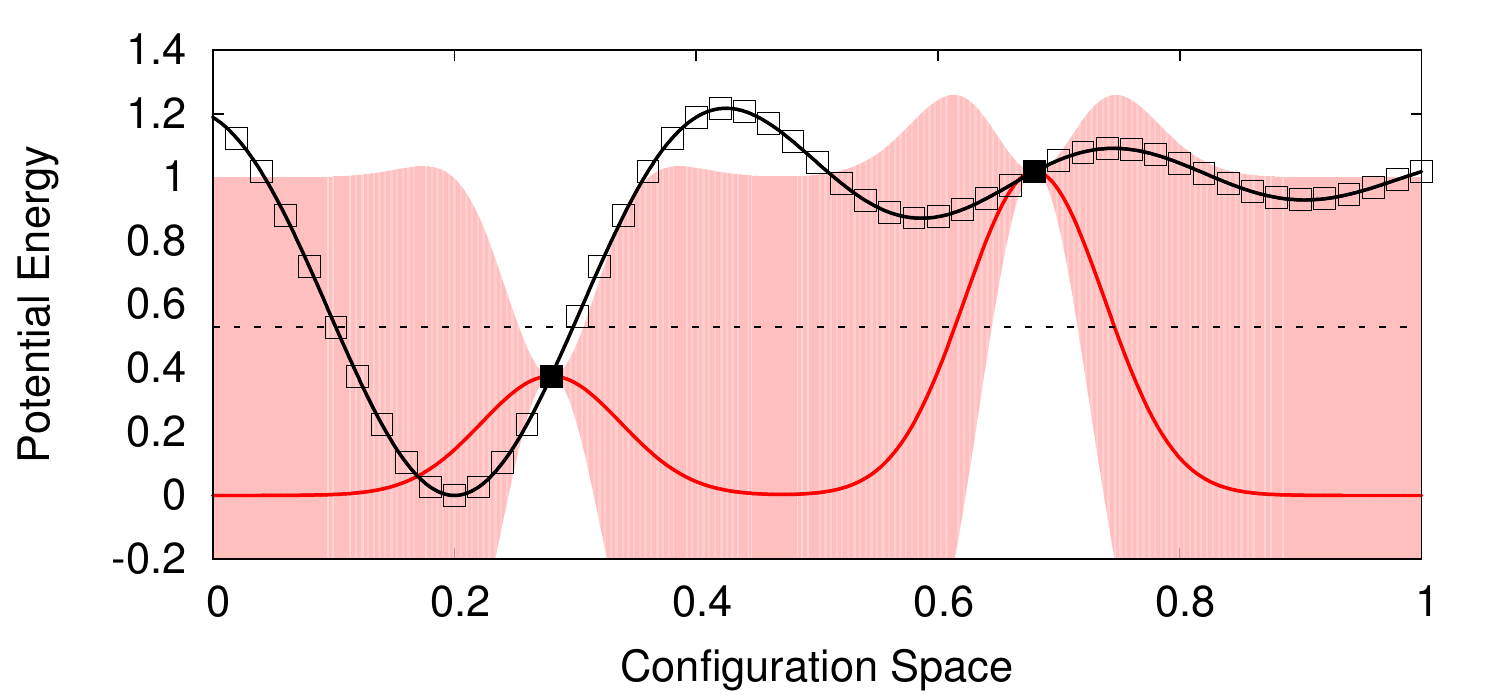} &
   \includegraphics[width=0.33\textwidth]{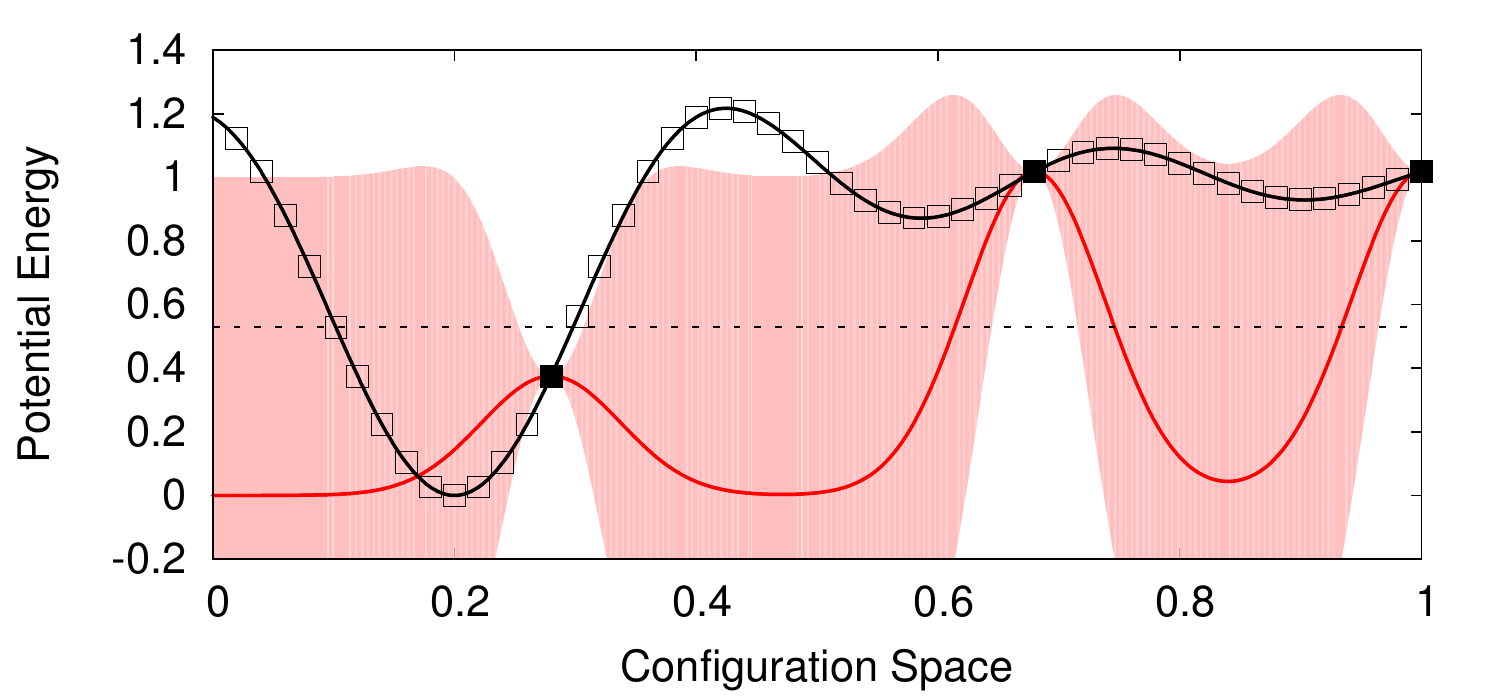} \\
   (a) low PE region &
   (b) step0 (initialization) &
   (c) step1 \\
   \includegraphics[width=0.33\textwidth]{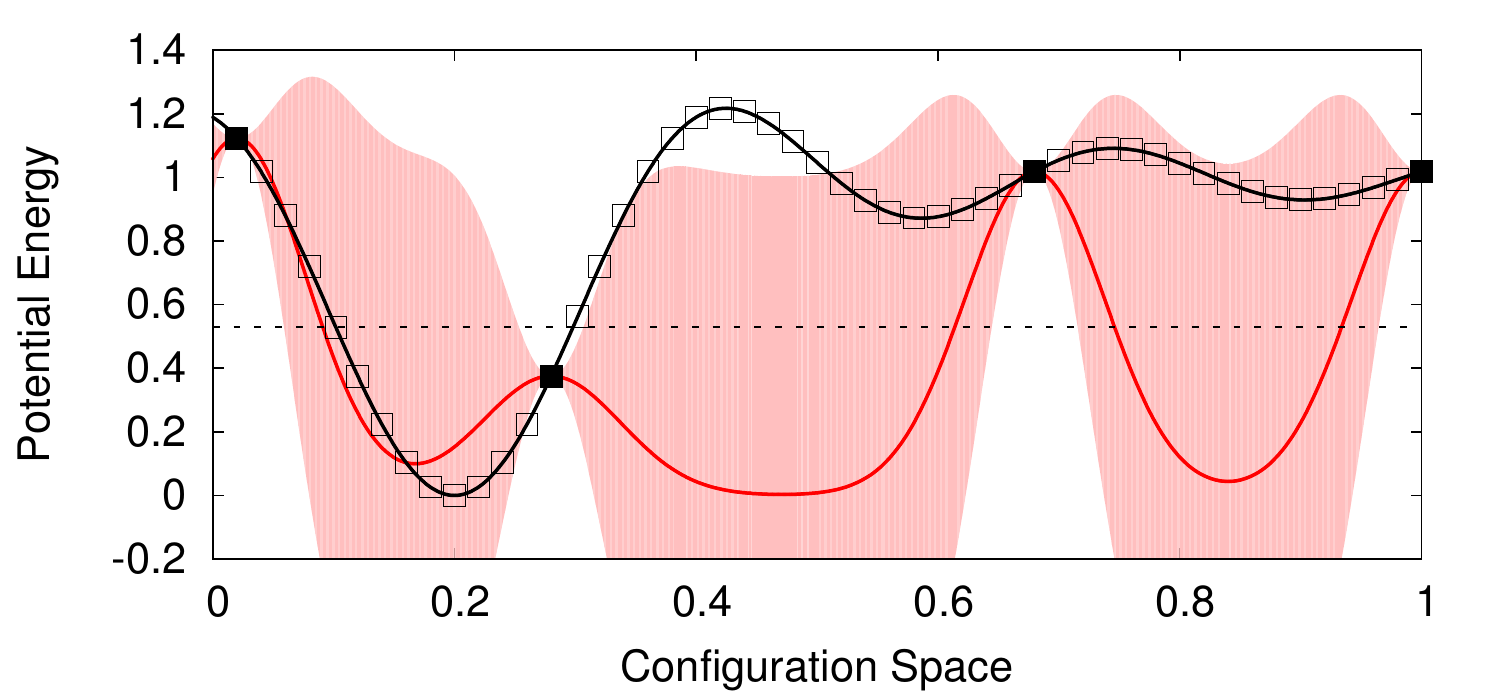} &
   \includegraphics[width=0.33\textwidth]{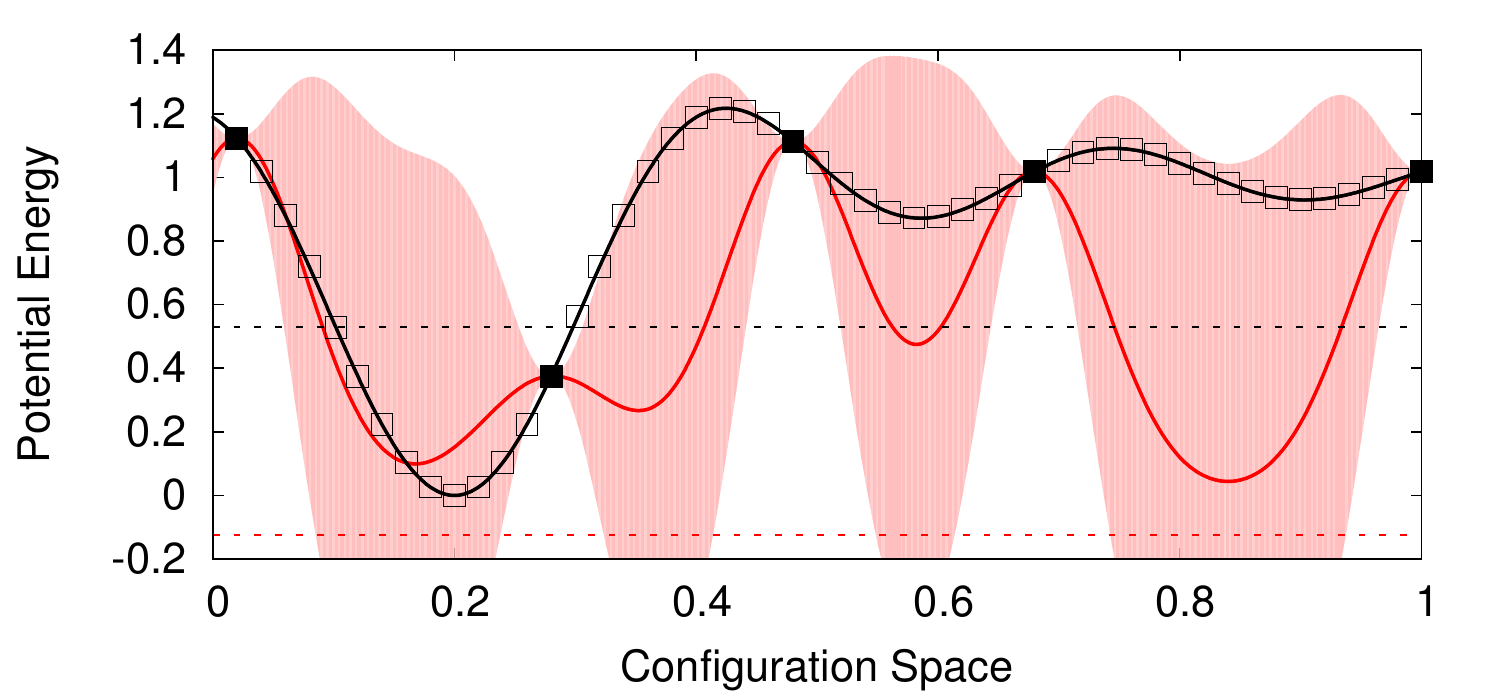} &
   \includegraphics[width=0.33\textwidth]{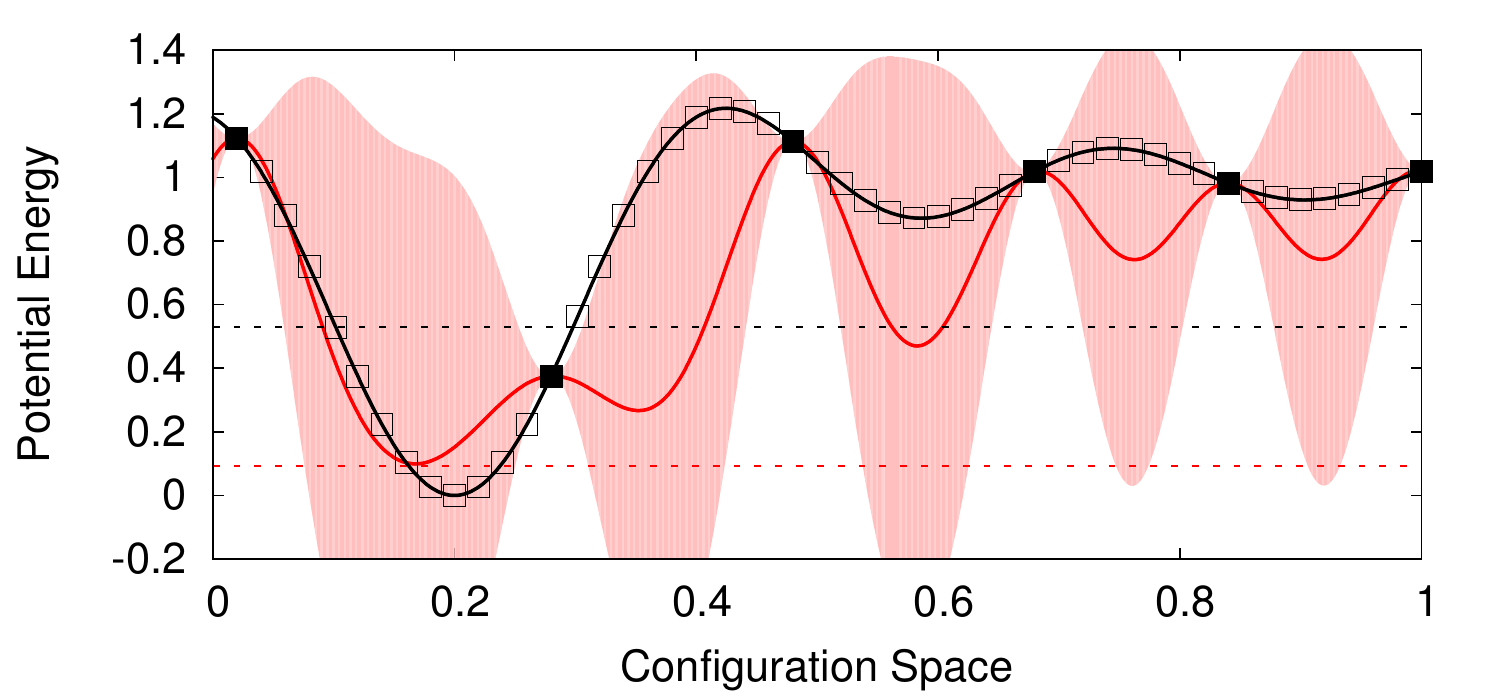} \\
   (d) step2 &
   (e) step3 &
   (f) step4 \\	   
   \includegraphics[width=0.33\textwidth]{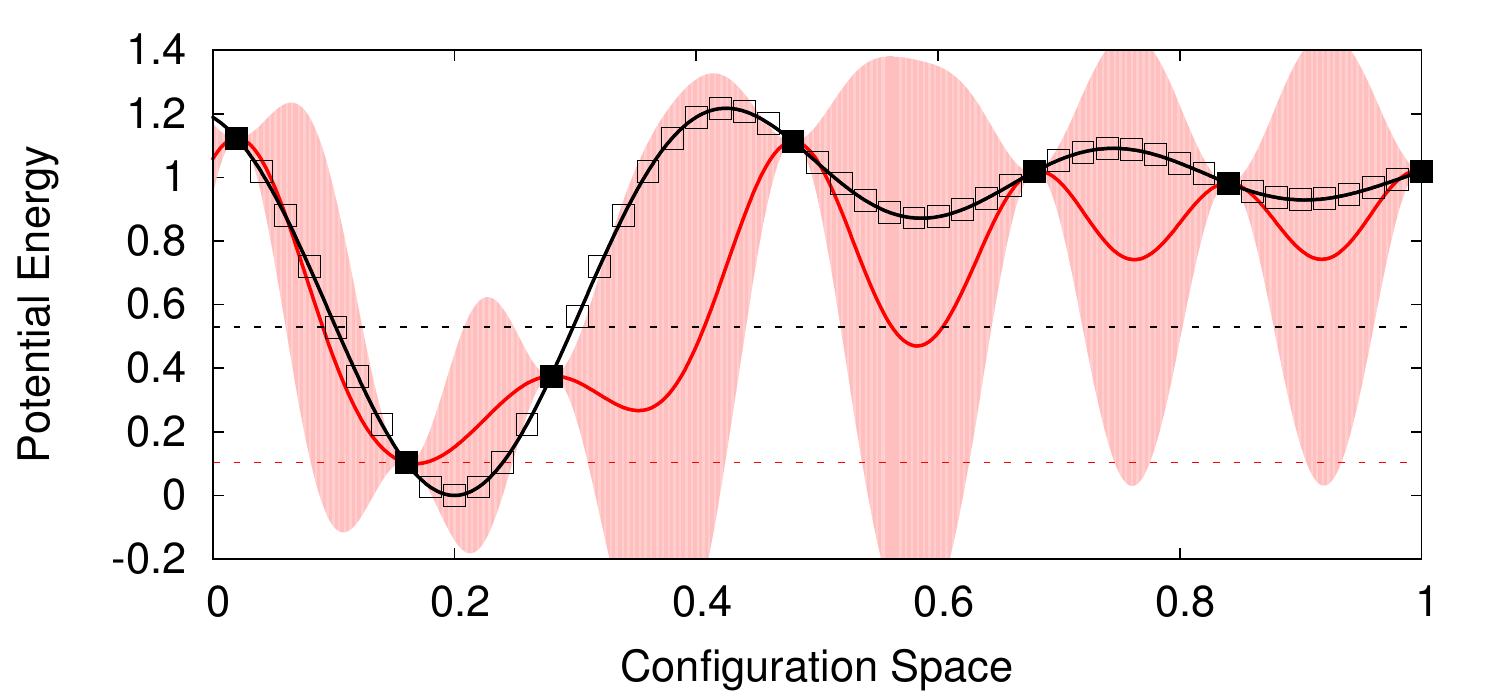} &
   \includegraphics[width=0.33\textwidth]{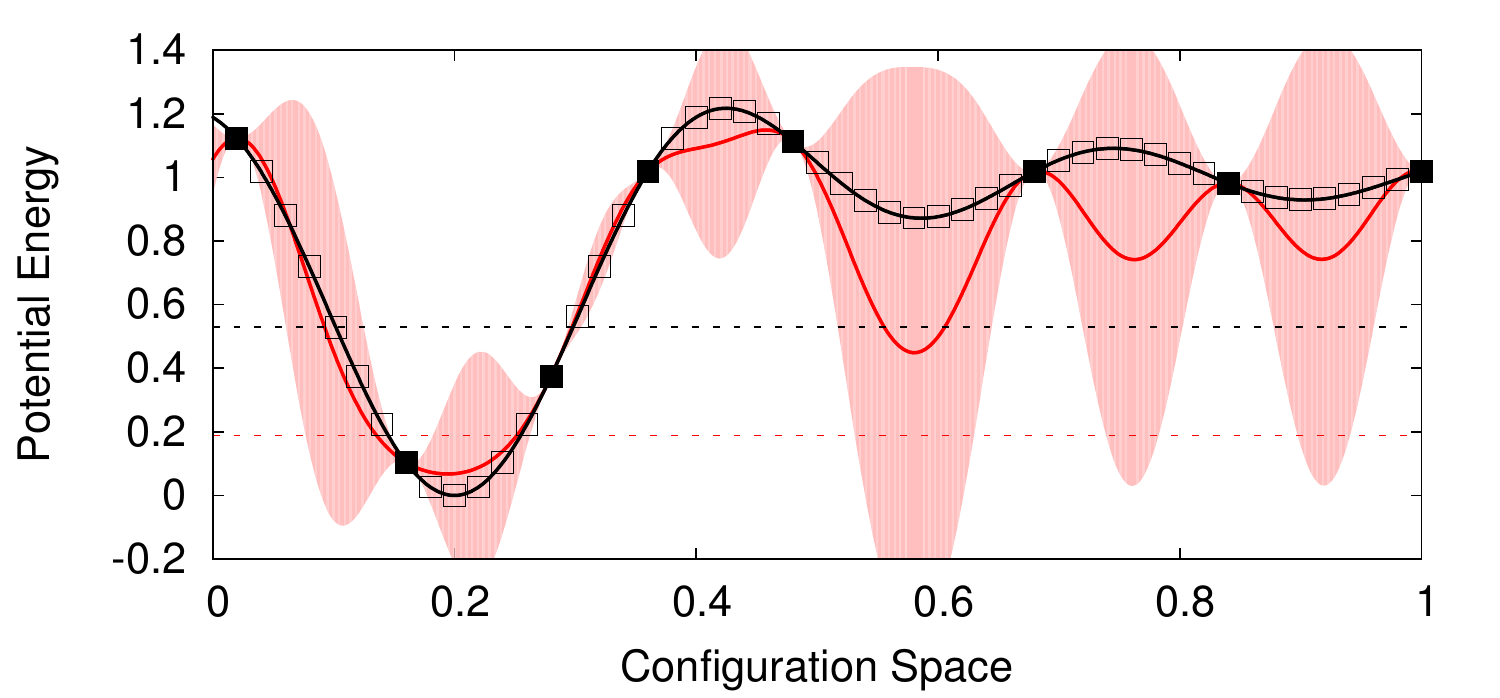} &
   \includegraphics[width=0.33\textwidth]{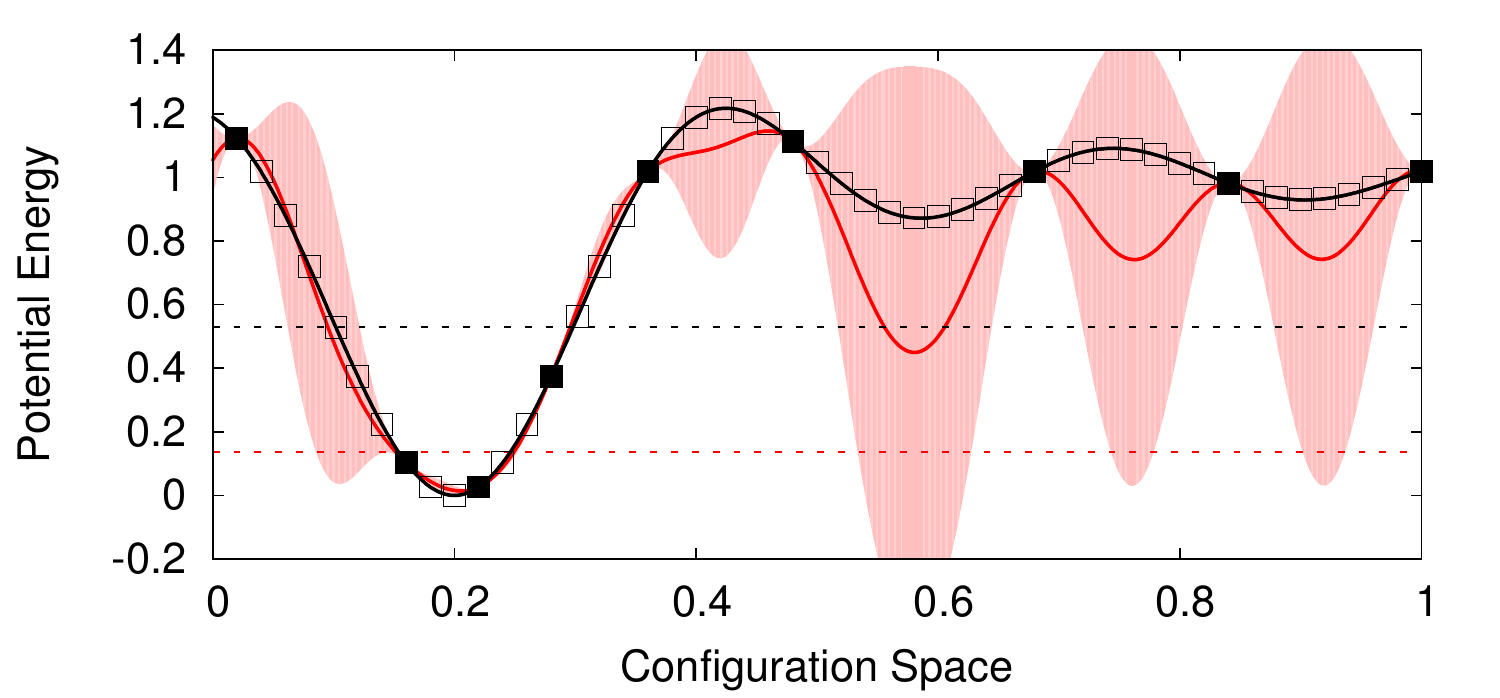} \\
   (g) step5 &
   (h) step6 &
   (i) step7 \\	   
   \includegraphics[width=0.33\textwidth]{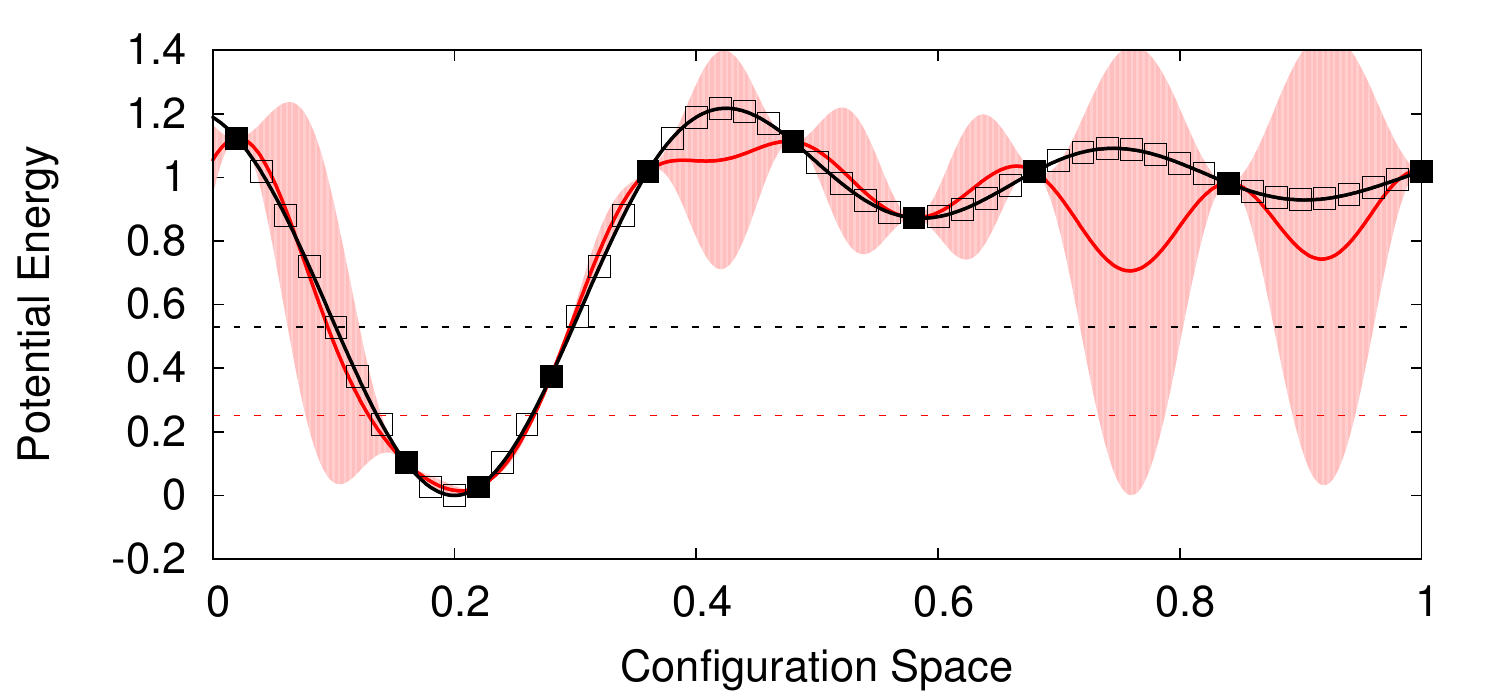} &
   \includegraphics[width=0.33\textwidth]{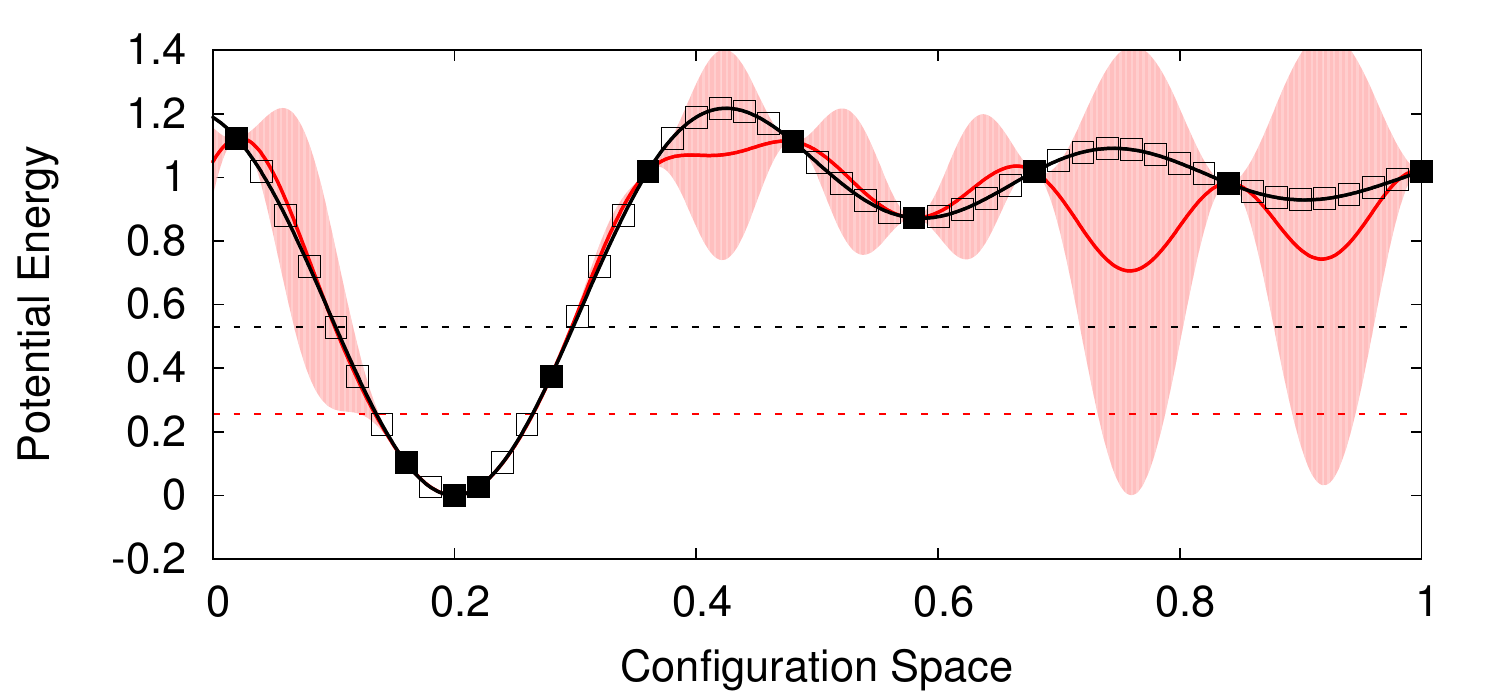} &
   \includegraphics[width=0.33\textwidth]{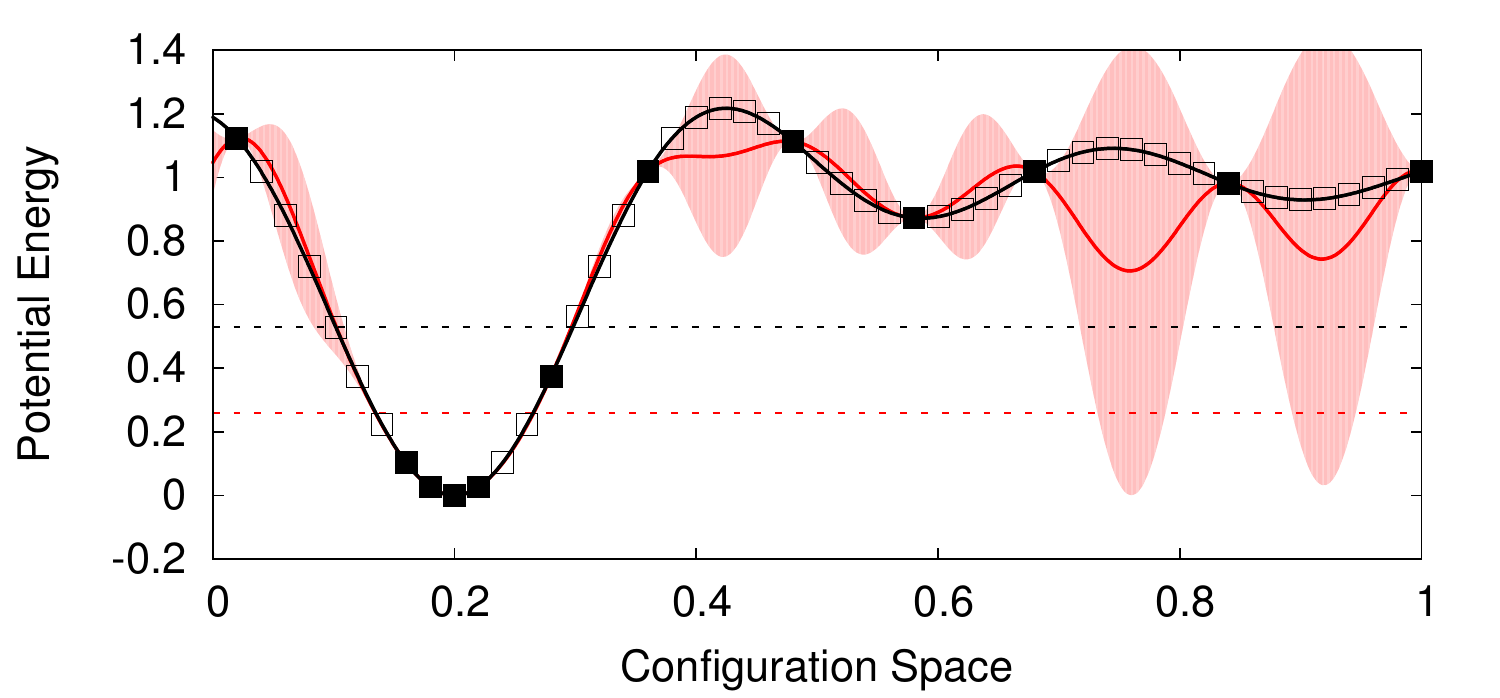} \\
   (j) step8 &
   (k) step9 &
   (l) step10 \\	   
   \includegraphics[width=0.33\textwidth]{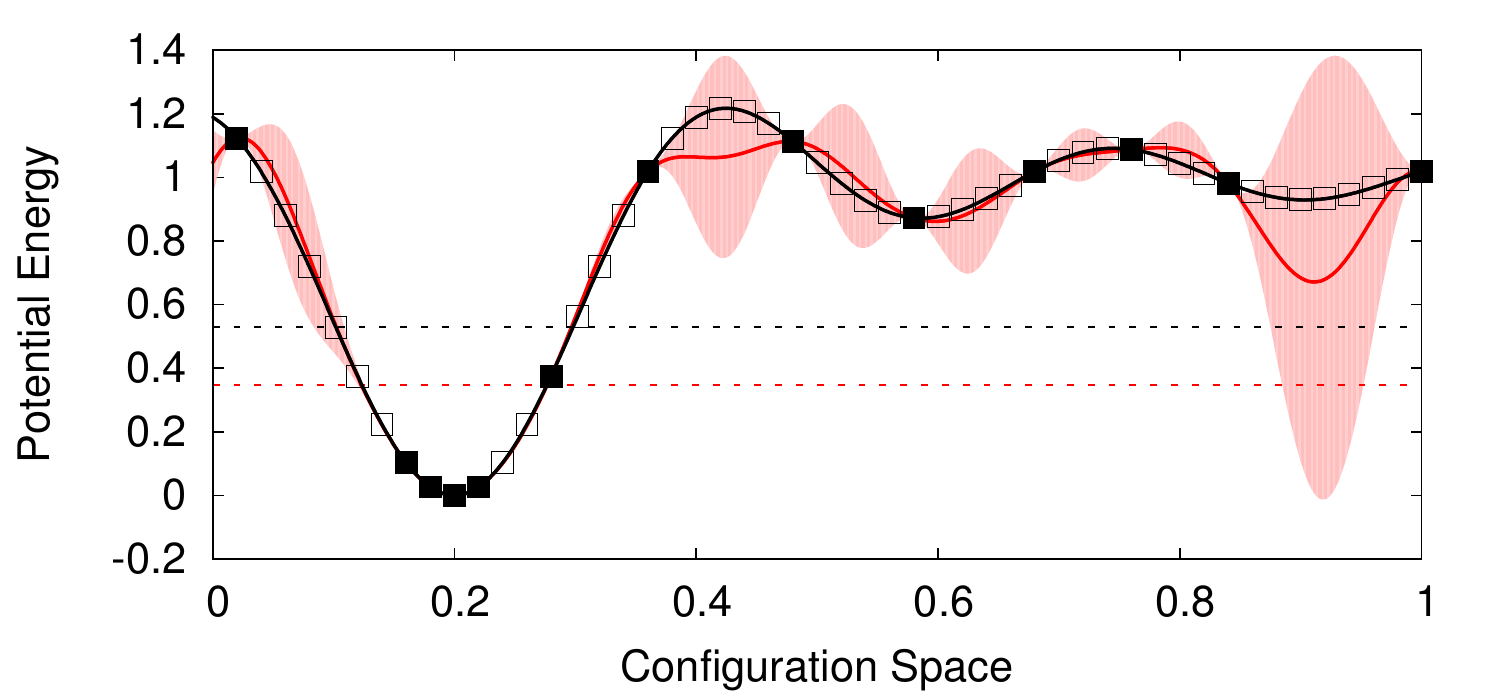} &
   \includegraphics[width=0.33\textwidth]{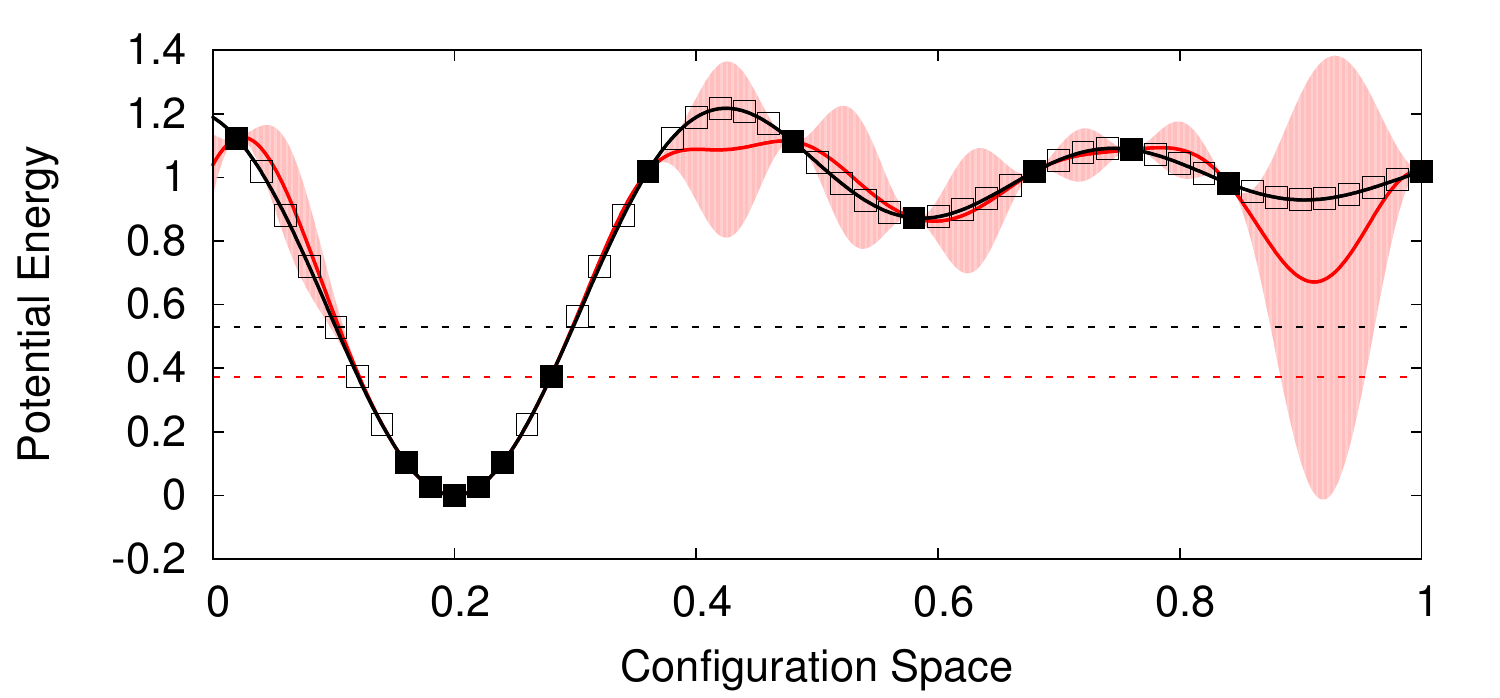} &
   \includegraphics[width=0.33\textwidth]{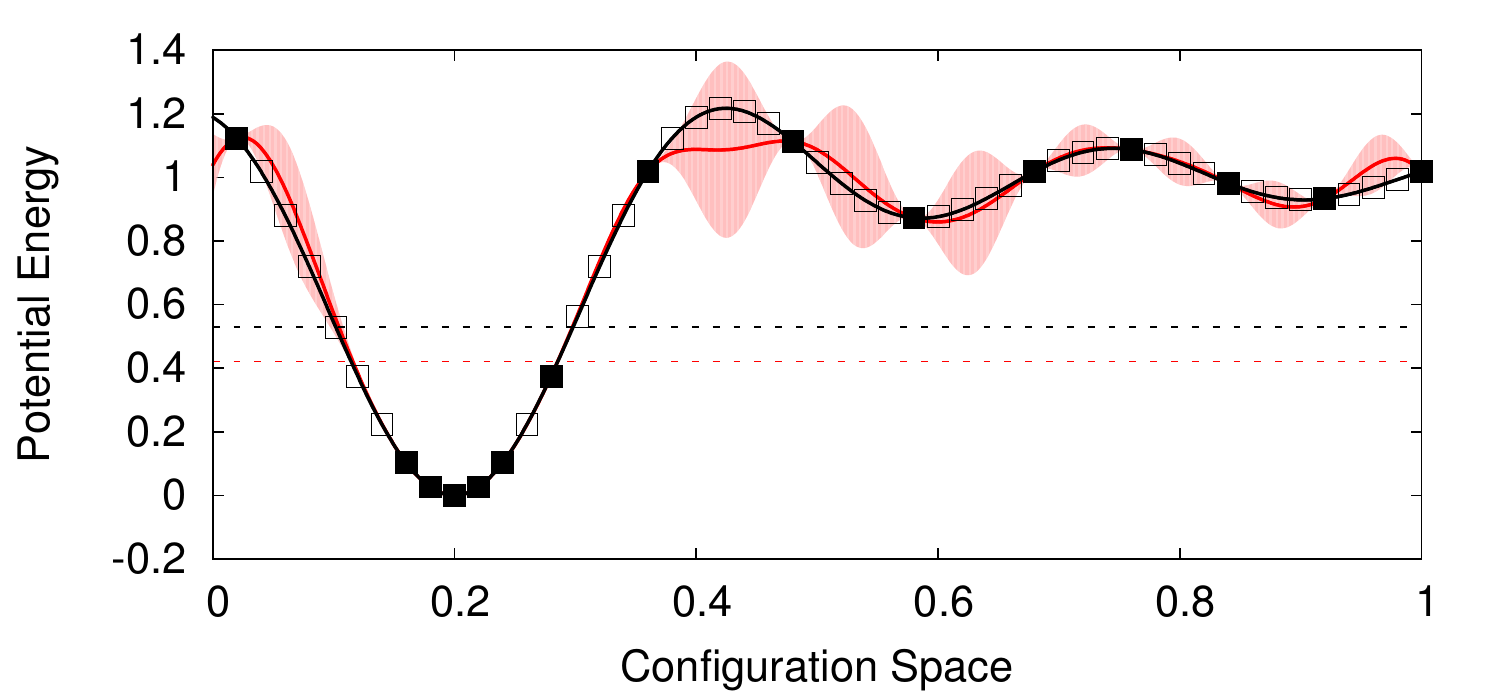} \\
   (m) step11 &
   (n) step12 &
   (o) step13 \\	   
   \includegraphics[width=0.33\textwidth]{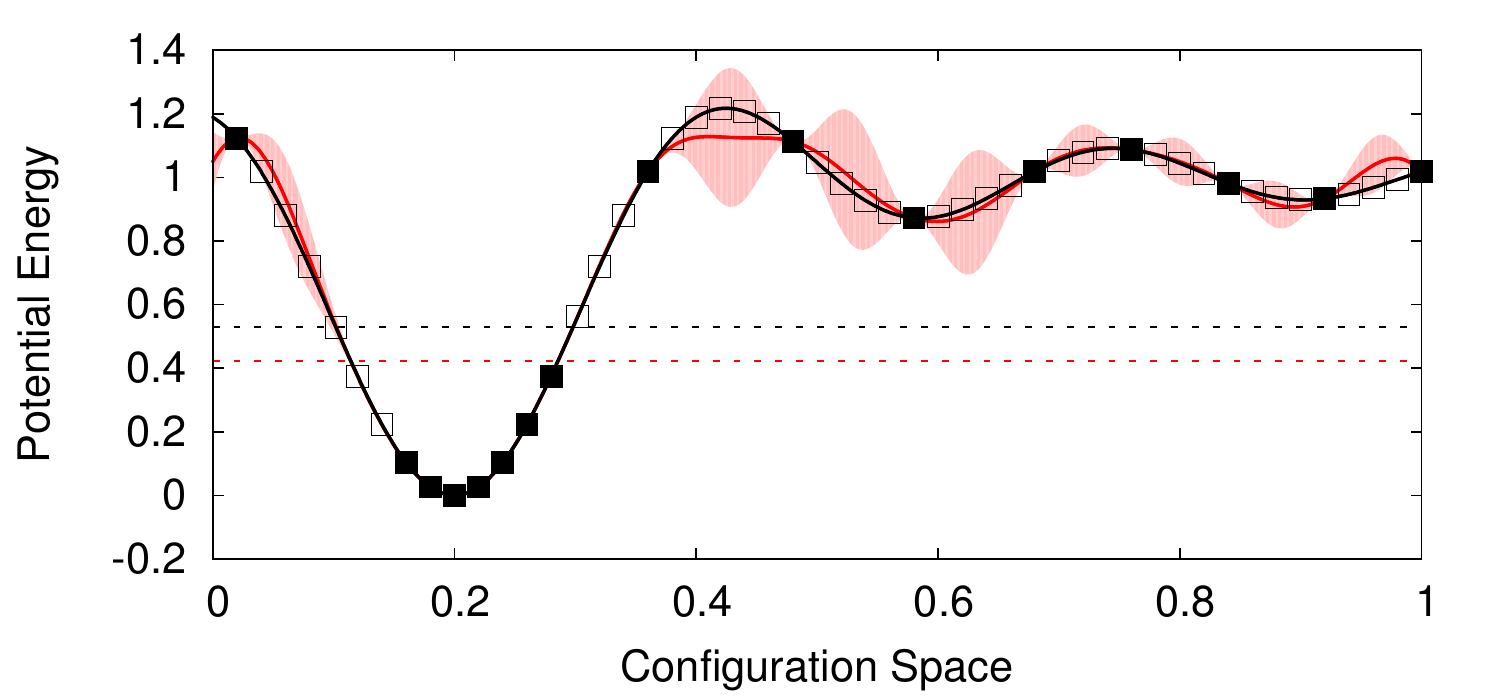} &
   \includegraphics[width=0.33\textwidth]{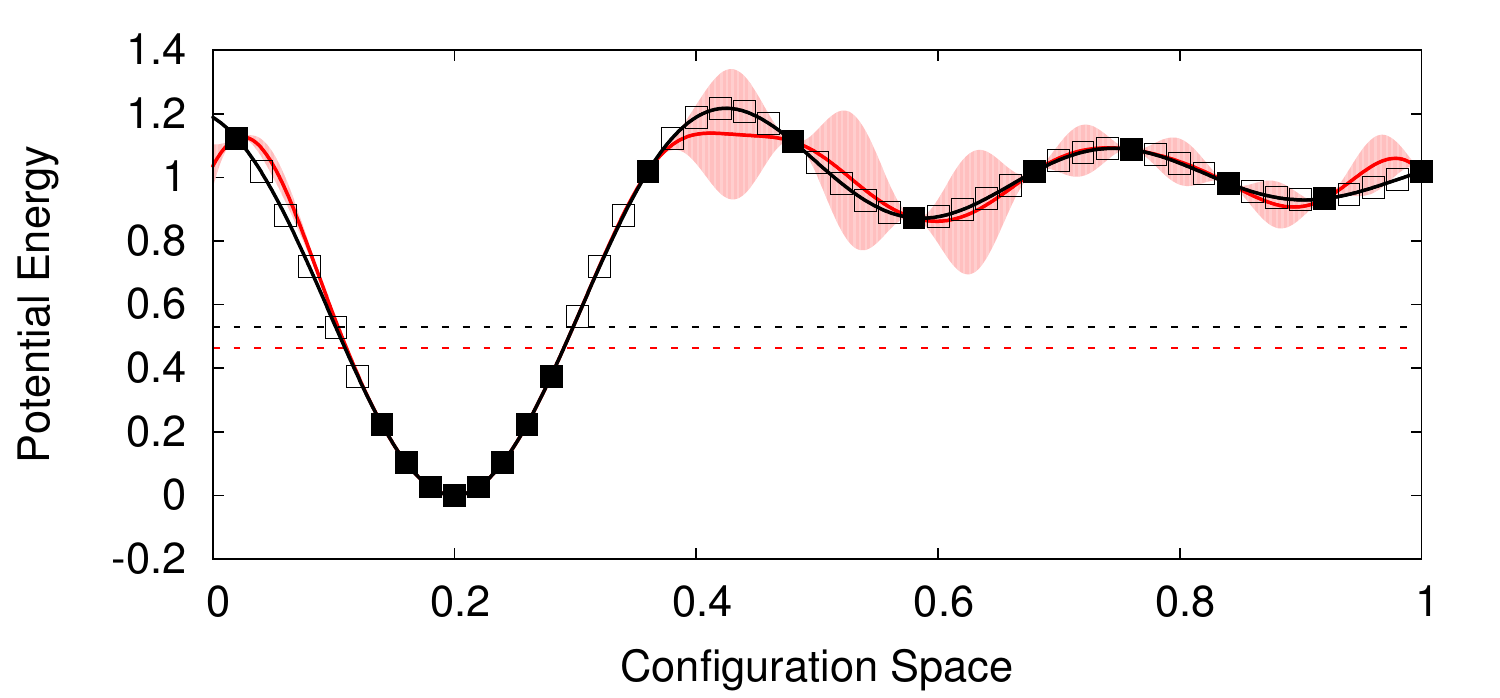} &
   \includegraphics[width=0.33\textwidth]{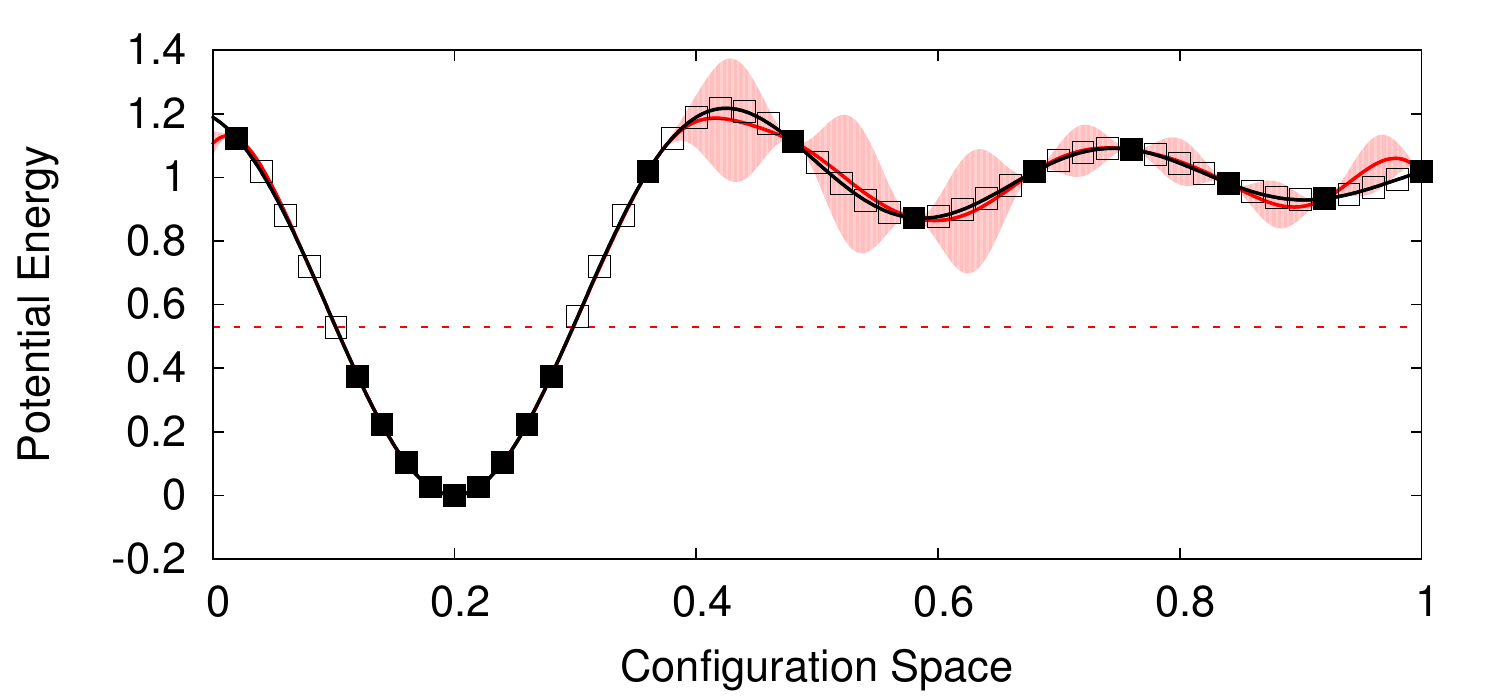} \\
   (p) step14 &
   (q) step15 &
   (r) step16 \\	   
  \end{tabular}
  \caption{
  Schematic illustration of the proposed selective sampling procedure 
  in one-dimensional configuration space with synthetic data. 
  In each plot, 
  the X and Y-axes represent 
  the configuration space and the PEs, respectively.
The PES and its $N=50$ points are shown as a 
black solid curve
and 
squares, respectively.
In plot (a), 
the low PE region with $\alpha = 0.20$ is represented by a blue bar.
The goal in this example is to efficiently identify and evaluate the PEs at the 9 points within this blue bar region. 
Plot (b) indicates the initialization step,
where two points (filled squares) are randomly selected and their PEs are evaluated.
The remaining 16 plots (plots (c) to plot (r))
indicate steps 1 to 16 of the procedure. 
At each step,
a point is selected and the PE is evaluated
by using the GP-based statistical model,
where
the points selected so far are shown as filled squares. 
In each plot, 
the predicted PEs and their uncertainties
given by the GP-based statistical model are depicted by 
a red solid curve and shaded region,
respectively. 
In addition,
the true and the estimated threshold values
$\theta_\alpha$
and 
$\hat{\theta}_\alpha$
are shown as black and red broken lines,
respectively. 
In this example, 
the 9 points in the low PE region are successfully identified
and the threshold is also correctly estimated
at step 16 (plot (r)).
  }
  \label{fig:GPSampling-Example}
 \end{center}
\end{figure*}

These features are made possible by exploiting an advantage of 
GP that it provides not only the predicted PE value
but also the uncertainty at each grid point.
Using this property of GP modeling,
we develop a new sampling procedure 
which
enables us to sample grid points 
preferentially 
from the low PE region, 
and 
tells us when we should stop sampling. 
\figurename \ref{fig:GPSampling-Example} illustrates the sequences of
the selective sampling by taking a one-dimensional synthetic PES as an
example, in which 9 grid points in the low PE region (within the blue
bar) should be selectively sampled from all the points (50 points).
Roughly speaking,
at each step, 
a point most likely to be located in the low PE region is sampled
in the light of the predicted PEs (red solid curve) and the
uncertainties (pale red area).
In early steps,
the predicted PEs are quite uncertain and have large discrepancies
from the true PES (black solid curve),
resulting in selecting grid points whose uncertainties are large. 
%
%
As the sampling proceeds, the predicted PE curve gradually approaches the true one with decreasing uncertainty, so that grid points in the low PE region are selectively sampled in later steps. The uncertainty in the GP model is useful also in the judgment to stop the sampling. The stopping criterion should be determined on the basis of the existence probability of unsampled low PE points, for which the information on the uncertainty is indispensable in addition to the predicted PEs.

In the remaining part of this section, we first present the problem setup in \S\ref{subsec:problem-setup}, and then describe the details of the three key features in \S\ref{subsec:key-feature1}, \S\ref{subsec:key-feature2}, and \S\ref{subsec:key-feature3}.

\subsection{Problem setup}
\label{subsec:problem-setup}

Suppose that there are $N$ grid points, $i = 1, \ldots, N$, in the configuration space. 
For each point, we denote by $E_i$ the potential energy (PE).
In order to investigate a target physical property such as proton conductivity,
we need to identify a region whose PEs are
\emph{relatively}
low.
Using a parameter 
$0 < \alpha < 1$, 
we define such a \emph{relatively} low PE region 
as the set of
$\alpha N$
points
at which the PEs
are lower than the PEs at the other 
$(1-\alpha)N$
points.
We refer to such a subset of points as
the ``low PE region'',
and the remaining set of points as
the ``high PE region''.
The goal of the problem is to identify all the points
in the low PE region
as efficiently as possible. 
For simplicity,
we assume that $\alpha$ is pre-specified,
but it can be adaptively determined 
as we demonstrate in 
\S~\ref{subsec:select-alpha}.

The problem of identifying the low PE region is formulated as follows.
Let $\theta_\alpha$ represent the threshold of the low PE region, 
and define 
\begin{align*}
 \cP_\alpha &:= \{i \in \{1, \ldots, N\} ~|~ E_i < \theta_\alpha\}, \\
 \cN_\alpha &:= \{i \in \{1, \ldots, N\} ~|~ E_i \ge \theta_\alpha\}.
\end{align*}
Then, the task is formally stated as the problem of identifying
all the points in $\cP_\alpha$.
Borrowing the terminology in statistics,
we refer to the points in $\cP_\alpha$ and $\cN_\alpha$ as \emph{positive} and \emph{negative} points, respectively. 

Note that we do not know
$\cP_\alpha$,
$\cN_\alpha$ 
and
$\theta_\alpha$
unless we actually compute the PEs at all the grid points. 
Therefore,
during the sampling process, 
these quantities are estimated based on the PEs at the points sampled in earlier steps. 
We denote our estimates of positive and negative sets as $\hat{\cP}_\alpha$ and $\hat{\cN}_\alpha$, respectively,
i.e.,
the former indicates the set of points at which the PEs have already been sampled and evaluated in earlier steps,
while the latter indicates the set of points at which the PEs have not yet been evaluated.
The proposed selective sampling procedure can be interpreted as the process of sequentially updating these two sets of points.
Specifically,
we start from $\hat{\cP}_\alpha = \emptyset$ and $\hat{\cN}_\alpha = \{1, \ldots, N\}$,
and then update the two sets as
$\hat{\cP}_\alpha \lA \hat{\cP}_\alpha \cup \{i^\prime\}$
and
$\hat{\cN}_\alpha \lA \hat{\cN}_\alpha \setminus \{i^\prime\}$
where
$i^\prime$
is the sampled point in the step.
When the stopping criterion is satisfied,
$\hat{\cP}_\alpha$
is expected to contain all the points in $\cP_\alpha$ with high probability.
We also denote the estimate of
$\theta_\alpha$
as
$\hat{\theta}_\alpha$
(see \S\ref{subsec:key-feature2} for how to estimate $\theta_\alpha$). 

\subsection{Gaussian process (GP) models}
\label{subsec:key-feature1}
We adopt GP
\cite{williams2006gaussian,stein2012interpolation}
as a choice of the statistical model of the PES.
Using a GP model, 
the potential energy
$E_i$ 
is represented in the following form: 
\begin{align}
 \label{eq:PE-model}
 E_i ~\sim~ N(\mu_i, \sigma^2_i),
 ~
 i = 1, \ldots, N, 
\end{align}
where
$N(\mu_i, \sigma^2_i)$
denotes the Normal distribution 
with mean 
$\mu_i$ 
and variance
$\sigma^2_i$. 

A GP model is a sort of regression model.
Let us consider
a $d$-dimensional vector of descriptors
for each point 
and denote the vector as
$\bm \chi_i \in \RR^d$
for
$i = 1, \ldots, N$. 
Then,
the mean and variance of the PE at the $i^{\rm th}$ point
are represented as functions of $\bm \chi_i$
as in
\eq{eq:m-func}
and 
\eq{eq:v-func}
below.
In a GP model, 
a so-called \emph{kernel function}
$k: \RR^d \times \RR^d \to \RR$
is employed. 
For two different points indexed by $i$ and $j$, 
$k(\bm \chi_i, \bm \chi_j)$ 
is roughly interpreted as a similarity between these two points. 
One of the most commonly used kernel functions is the \emph{RBF kernel}
\begin{align}
 \label{eq:rbf-kernel}
 k(\bm \chi, \bm \chi^\prime) = \gamma_1^2 \exp(- \| \bm \chi - \bm \chi^\prime\|^2 / 2 \gamma_2^2),
\end{align}
where
$\gamma_1, \gamma_2 > 0$
are tuning parameters,
and
$\| \cdot \|$
represents the $L_2$ norm.
Furthermore,
for $n$ points
indexed by $1, \ldots, n$, 
let
$\bm K \in \RR^{n \times n}$
be the so-called \emph{kernel matrix} defined as
\begin{align}
 \bm K := \mtx{ccc}{
 k(\bm \chi_1, \bm \chi_1) &
 \cdots &
 k(\bm \chi_1, \bm \chi_n) \\
 \vdots &
 \ddots &
 \vdots \\
 k(\bm \chi_n, \bm \chi_1) &
 \cdots &
 k(\bm \chi_n, \bm \chi_n)}.
\end{align}
Then,
for any points in the configuration space whose descriptor vector is represented as 
$\bm \chi \in \RR^d$,
the GP model provides the predictive distribution of its PE 
in the form of a Normal distribution
$N(\mu(\bm \chi), \sigma^2(\bm \chi))$.
Here, 
the mean function $\mu: \RR^d \to \RR$ is given as 
\begin{align}
 \label{eq:m-func}
 \mu(\bm \chi) := \bm \kappa(\bm \chi)^\top \bm K^{-1} \bm E,
\end{align}
where
$\bm \kappa(\bm \chi) := [k(\bm \chi, \bm \chi_1), \ldots, k(\bm \chi, \bm \chi_n)]^\top$
and
$\bm E := [E_1, \ldots, E_n]^\top$,
while 
the variance function
$\sigma^2: \RR^d \to \RR$
is given as
\begin{align}
 \label{eq:v-func}
 \sigma^2(\bm \chi) := k(\bm \chi, \bm \chi) - \bm \kappa(\bm \chi)^\top \bm K^{-1} \bm \kappa(\bm \chi).
\end{align}
At each step of the procedure, 
we fit a GP model of PES 
based on
$\{(\bm \chi_i, E_i)\}_{i \in \hat{\cP}_\alpha}$,
which is 
the set of points that have been selected and whose PEs have been already evaluated
by DFT calculations 
in earlier steps. 

\subsection{Selection criterion for low PE region identification}
\label{subsec:key-feature2}

Given a GP model
in the form of \eq{eq:PE-model}
for each point,
the subsequent task is to select the point
at which the PE is most likely to be lower
than the estimated threshold
$\hat{\theta}_\alpha$
(we will discuss how to estimate the threshold later). 
In this task,
we can borrow some techniques developed in the context of
\emph{Bayesian optimization}
\cite{mockus1994application,brochu2010tutorial},
which has been used for minimization or maximization of an unknown function.
In the Bayesian optimization literature,
there are two main options that can be adapted to our task.
The first option is to select the point
at which the probability that the PE is lower than $\hat{\theta}_\alpha$ is maximized,
which is called \emph{probability of improvement},
and formulated as
\begin{align}
 \label{eq:probability-improvement}
 i^\prime 
 :=
 \arg \max_{i \in \hat{\cN}_\alpha}
 ~
 \Phi(\hat{\theta}_\alpha ; \mu(\bm \chi_i), \sigma^2(\bm \chi_i)),
\end{align}
where
$\Phi(\cdot ; \mu, \sigma^2)$
is the cumulative distribution function of
$N(\mu, \sigma^2)$.
The second option is
\emph{expected improvement},
which is similarly formulated as 
\begin{align}
 \label{eq:expected-improvement}
 i^\prime 
 :=
 \arg \min_{i \in \hat{\cN}_\alpha}
 ~
 \int_{-\infty}^{\hat{\theta}_\alpha}
 E ~ \phi(E; \mu(\bm \chi_i), \sigma^2(\bm \chi_i)) dE, 
\end{align}
where
$\phi(\cdot ; \mu, \sigma^2)$
is the probability density function of
$N(\mu, \sigma^2)$.
In the simulation studies conducted in \S\ref{sec:result},
we used \eq{eq:expected-improvement}. 
In our experience, 
there is little difference in the performances 
between the choices of 
\eq{eq:probability-improvement}
and 
\eq{eq:expected-improvement}. 

In order to obtain an estimate $\hat{\theta}_\alpha$ of the threshold $\theta_\alpha$, 
let us consider the following contingency table
\begin{align}
 \label{eq:contingency-table}
 \begin{tabular}{l|c|c}
  & $\cP_\alpha$ & $\cN_\alpha$ \\ \hline
  $\hat{\cP}_\alpha$ & \#TP$(\hat{\theta}_\alpha)$ & \#FP$(\hat{\theta}_\alpha)$ \\ \hline
  $\hat{\cN}_\alpha$ & \#FN$(\hat{\theta}_\alpha)$ & \#TN$(\hat{\theta}_\alpha)$
 \end{tabular}
\end{align}
where
TP, FP, FN, and TN
stand for true positive, false positive, false negative, and true negative, respectively,
and the notation  $\#$ indicates the number of the event. 
These numbers for four events can be rephrased as 
\begin{itemize}
\item \#TP: The number of sampled points in the low PE region.
\item \#FP: The number of sampled points in the high PE region.
\item \#FN: The number of not-yet sampled points in the low PE region.
\item \#TN: The number of not-yet sampled points in the high PE region.
\end{itemize}
Note that,
in \eq{eq:contingency-table},
these numbers depend on $\hat{\theta}_\alpha$.
Remembering that
$\cP_\alpha / (\cP_\alpha + \cN_\alpha) = \alpha$,
the following relationship should be maintained
\begin{align}
 \label{eq:requirement}
 \frac{
 \#{\rm TP}(\hat{\theta}_\alpha) +
 \#{\rm FN}(\hat{\theta}_\alpha)
 }{
 N
 }
 = \alpha.
\end{align}
Since we have already evaluated
$E_i$ for $i \in \hat{\cP}_\alpha$,
we can simply obtain
\begin{align}
 \label{eq:TP}
 \#{\rm TP}(\hat{\theta}_\alpha) &= \sum_{i \in \hat{\cP}_\alpha} I(E_i < \hat{\theta}_\alpha),
\end{align}
where $I(\cdot)$ is the indicator function, defined by 
$I(z) = 1$
if $z$ is true 
and 
$I(z) = 0$
if $z$ is false.
On the other hand,
we need to estimate
$\#{\rm FN}(\hat{\theta}_\alpha)$ 
based on the statistical model $\eq{eq:PE-model}$
because we do not know
$E_i$ for $i \in \hat{\cN}_\alpha$:
\begin{align}
 \label{eq:FN}
 \hspace*{-2.5mm}
 \#{\rm FN}(\hat{\theta}_\alpha)
 \simeq
 \#{\hat{\rm FN}}(\hat{\theta}_\alpha)
 :=
 \sum_{i \in \cN_\alpha} \Phi(\hat{\theta}_\alpha ; \mu(\bm \chi_i), \sigma^2(\bm \chi_i)).
\end{align}
The estimate of the threshold $\hat{\theta}_\alpha$ is determined in each step
so that it satisfies the requirement \eq{eq:requirement}
where the quantities in the left-hand side are given by
\eq{eq:TP} and \eq{eq:FN}.

\subsection{How to assess the sampled points}
\label{subsec:key-feature3}

When the sampling is stopped,
$\hat{\cP}_\alpha$
should ideally contain all the points in
$\cP_\alpha$,
i.e., $\hat{\cP}_\alpha \supseteq \cP_\alpha$.
As we can easily notice from the contingency table in
\eq{eq:contingency-table},
this requirement can be rewritten as
$\#{\rm FN}(\hat{\theta}_\alpha) = 0$.
This indicates that
the estimated false negative rate (FNR) defined as 
\begin{align}
 \label{eq:FNR}
 \hat{\rm FNR} := \frac{\#{\hat{\rm FN}}(\hat{\theta}_\alpha)}{\#{\rm TP}(\hat{\theta}_\alpha) + \#{\hat{\rm FN}}(\hat{\theta}_\alpha)}
\end{align}
can be used for assessing the \emph{badness} of the sampled points.
$\hat{\rm FNR}$ in \eq{eq:FNR} can be interpreted as
the proportion of points for which the PEs have yet to be evaluated.
At each step,  we compute
$\#{\rm TP}(\hat{\theta}_\alpha)$
by \eq{eq:TP}
and
estimate
$\#{\rm FN}(\hat{\theta}_\alpha)$
by \eq{eq:FN}.
Then, the sampling is terminated if $\hat{\rm FNR}$ is close to zero (e.g., $10^{-6}$).

 \section{Proton PES in barium zirconate}
 \label{sec:pesBaZrO3}

 \subsection{Grid points}
\begin{figure}[h]
\begin{center}
\begin{tabular}{cc}
\includegraphics[width=.22\textwidth]{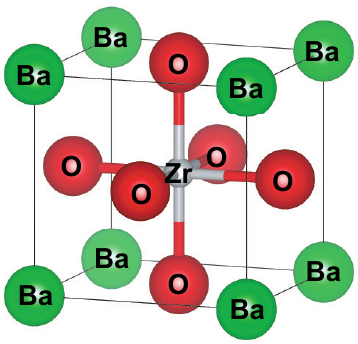} &
\includegraphics[width=.185\textwidth]{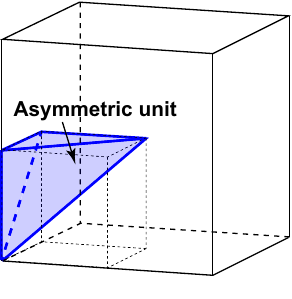} \\
(a) & (b)
\end{tabular}
\caption{
(a) Crystal structure of ${\rm BaZrO_3}$. 
(b) Asymmetric unit in the unit cell of ${\rm BaZrO_3}$.
}
\label{fig:BaZrO3}
\end{center}  
\end{figure}

To show the performance of the selective sampling procedure in the present study, we evaluated the PEs for all the grid points, i.e., the entire PES using DFT calculations with structure optimization. 
Firstly, grid points are introduced in the host crystal lattice of ${\rm BaZrO_3}$.
\figurename \ref{fig:BaZrO3} shows the crystal structure of ${\rm BaZrO_3}$ (space group: $Pm\overline{3}m$ (221)) and its asymmetric unit satisfying $0 \le x, y, z \le 0.5$, $y \le x$ and $z \le y$ where $x$, $y$ and $z$ denote the three-dimensional fractional coordinates of a proton introduced into the host lattice.
The lattice constant $a$ was experimentally reported to be 4.19-4.20 \AA\ \cite{pagnier2000neutron,levin2003phase}, which is reasonably consistent with our calculated value (4.23 \AA) using the computational condition as will be described later. 
${\rm Ba}$, ${\rm Zr}$ and ${\rm O}$ ions occupy $1a$, $1b$, and $3c$ sites, respectively, in the case of the origin setting shown in \figurename \ref{fig:BaZrO3}.

Considering a $40 \times 40 \times 40$ grid in the unit cell (grid interval is nearly equal to 0.1 \AA), there are 64,000 grid points in total. 
Reflecting the high crystallographic symmetry of ${\rm BaZrO_3}$, the number of grid points in the asymmetric unit shown in the right plot of \figurename \ref{fig:BaZrO3} is reduced to 1771 points. 
In other words, these PEs enable us to construct the entire three-dimensional PES using symmetry operations.
Among them, three points exactly coincide with a ${\rm Ba}$ ion, a ${\rm Zr}$ ion, and an ${\rm O}$ ion. Further removing those three points, the remaining grid points are reduced to $N = 1768$ points.

\subsection{Procedure of DFT calculation}

DFT calculations for the PES evaluation in ${\rm BaZrO_3}$ were based on the projector augmented wave (PAW) method as implemented in the VASP code \cite{blochl1994projector,kresse1993ab,kresse1996efficiency,kresse1999ultrasoft} The generalized gradient approximation (GGA) parameterized by Perdew, Burke, and Ernzerhof was used for the exchange-correlation term \cite{perdew1996generalized}. The $5s$, $5p$ and $6s$ orbitals for barium, $4s$, $4p$, $5s$ and $4d$ for zirconium, $2s$ and $2p$ for oxygen, and $1s$ for hydrogen were treated as valence states. The plane-wave cutoff energy was set to be 400 eV. A supercell consisting of $3 \times 3 \times 3$ unit cells (135 atoms) was used, with a $2 \times 2 \times 2$ mesh for the $k$-point sampling, in which the atomic positions only in the limited region corresponding to $2 \times 2 \times 2$ unit cells around the introduced proton were optimized with the other atoms and the proton fixed. The atomic positions were optimized until the residual forces converged to less than 0.02 eV/\AA.

\subsection{Evaluated PES}

\begin{figure}[h]
\begin{center}
\includegraphics[width=.42\textwidth]{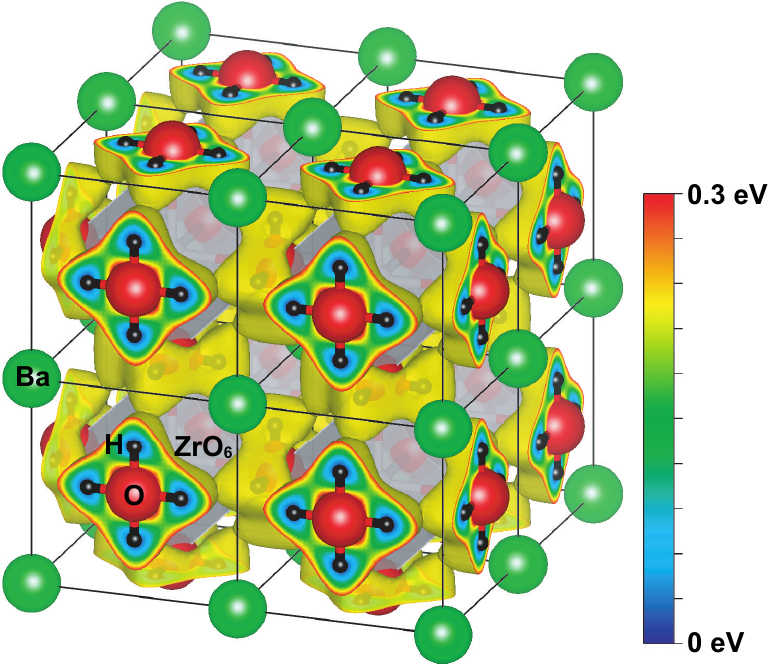}
\caption{PES of a proton in the low PE region ($\alpha = 0.2$) obtained by DFT calculations with structural optimization for ${\rm BaZrO_3}$. 
The low PE region ($\alpha = 0.2$) at which the PEs are lower than 0.30 eV is surrounded by yellow surfaces. }
\label{fig:PE-relaxed}
\end{center}  
\end{figure}

\figurename \ref{fig:PE-relaxed} shows the evaluated PES of a proton in a low PE region for ${\rm BaZrO_3}$.
All points in the low PE region with $\alpha = 0.2$ show a PE of less than 0.30 eV. 
The blue spheres bonding to single ${\rm O}$ ions in the figure denote the most stable proton sites, which are located $\sim 1$ \AA\ from the ${\rm O}$ ions.
The ${\rm OH}$ distance is equivalent to that in water, indicating that protons in ${\rm BaZrO_3}$ are stabilized by forming an OH bond. 
There are four equivalent proton sites per O ion, which are connected by low PE points around the O ion.
This is the rotational path around the O ion, whose potential barrier is 0.18 eV.
On the other hand, the hopping path connecting adjacent rotational orbits is located at the periphery of the edges of the $\rm ZrO_6$ octahedra. 
The calculated potential barrier of the hopping path is 0.25 eV, which is higher than that of the rotational path. 
The two kinds of paths form a three-dimensional proton-conducting network throughout the crystal lattice, meaning that protons migrate over a long range by repeated rotation and hopping. 
The conduction mechanism of the rotation and hopping has been previously studied based on DFT analysis with the nudged elastic band (NEB) method \cite{gomez2005effect,bjorketun2005kinetic,bork2010simple}. 
The reported potential barriers of the rotation and the hopping are in the range of 0.10-0.18 eV and 0.23-0.27 eV, respectively, which are in good agreement with our results.

\section{Results of selective samplings}
\label{sec:result}
In this section, 
we illustrate the advantage of the proposed selective sampling procedure
through a simulation study 
in which
we apply it 
to 
a low PE region identification problem 
for a proton in 
${\rm BaZrO_3}$. 

\subsection{Setup of the simulation study}
\label{subsec:}
We compared the performances of several sampling methods for $\alpha =$ 0.05, 0.1 and 0.2. 
Specifically,
we compared the following seven methods: 
\begin{itemize}
 \item {\tt GP1(xyz)}
 \item {\tt GP2(xyz + 1st NNs)}
 \item {\tt GP3(xyz + decaying NNs)}
 \item {\tt GP4(xyz + prePES)}
 \item {\tt random}
 \item {\tt prePES}       
 \item {\tt ideal}
\end{itemize}
The first four methods are
the proposed GP-based selective sampling procedure 
with different descriptors.
In {\tt GP1}, 
we used the three-dimensional coordinates
$x_i, y_i, z_i$
in the host crystal lattice 
as the descriptors of the $i^{\rm th}$ point
(denoted as ``{\tt xyz}''). 
In {\tt GP2},
we also used the first-nearest-neighbor (NN) distances
to
Ba, Zr and O
atoms 
from each point 
as additional descriptors 
(denoted as ``{\tt 1st NN}''). 
In {\tt GP3},
a weighted sum of the distances to the 1st NN atoms, 2nd NN atoms, and so on, 
are also incorporated
as additional descriptors, 
where the weights are decaying to zero as the atom goes away from the point 
(denoted as ``{\tt decaying NNs}''). 
In {\tt GP4},
we used
\emph{preliminary PES}
(denoted as ``{\tt prePES}'')
as an additional descriptor.
Here,
we mean by preliminary PES 
a \emph{rough but quick} approximation of the PES 
obtained by using less accurate but more efficient computational methods. 
Specifically, 
we used PE values at all the $N$ points
obtained by single-point DFT calculations
using a smaller supercell
consisting of $2 \times 2 \times 2$ unit cells
with a $2 \times 2 \times 2$ k-point sampling.
%
In {\tt GP4},
GP model is used for modeling the difference between
the actual PES
and
the preliminarily PES \cite{Ramakrishnan15}.

The method
{\tt random} indicates a naive random sampling,
where
a point is selected at each step 
uniformly at random. 
The method 
{\tt prePES} indicates a selective sampling method 
based only on a preliminary PES.
Specifically, 
points are sequentially selected
in the ascending order of the preliminary PEs
obtained by single-point DFT calculations. 
Finally,
{\tt ideal}
indicates the best possible ideal sampling method 
which is only possible
when the actual PEs at all the points are known in advance.

In {\tt GP1} to {\tt GP4}, we must select two points at random for initializing the GP model.
Thus, 
we report the average and the standard deviation over 10 runs with different random seeds.
The tuning parameters of the GP models were set as
$\gamma_1 = \gamma_2 = 0.5$.
According to our preliminary experiments (not shown),
the performances were not very sensitive to the choices of those tuning parameters.

\subsection{Performance comparison}
\label{subsec:performance-comparison}

\begin{figure*}[htbp]
\begin{center}
\includegraphics[width=0.75\textwidth]{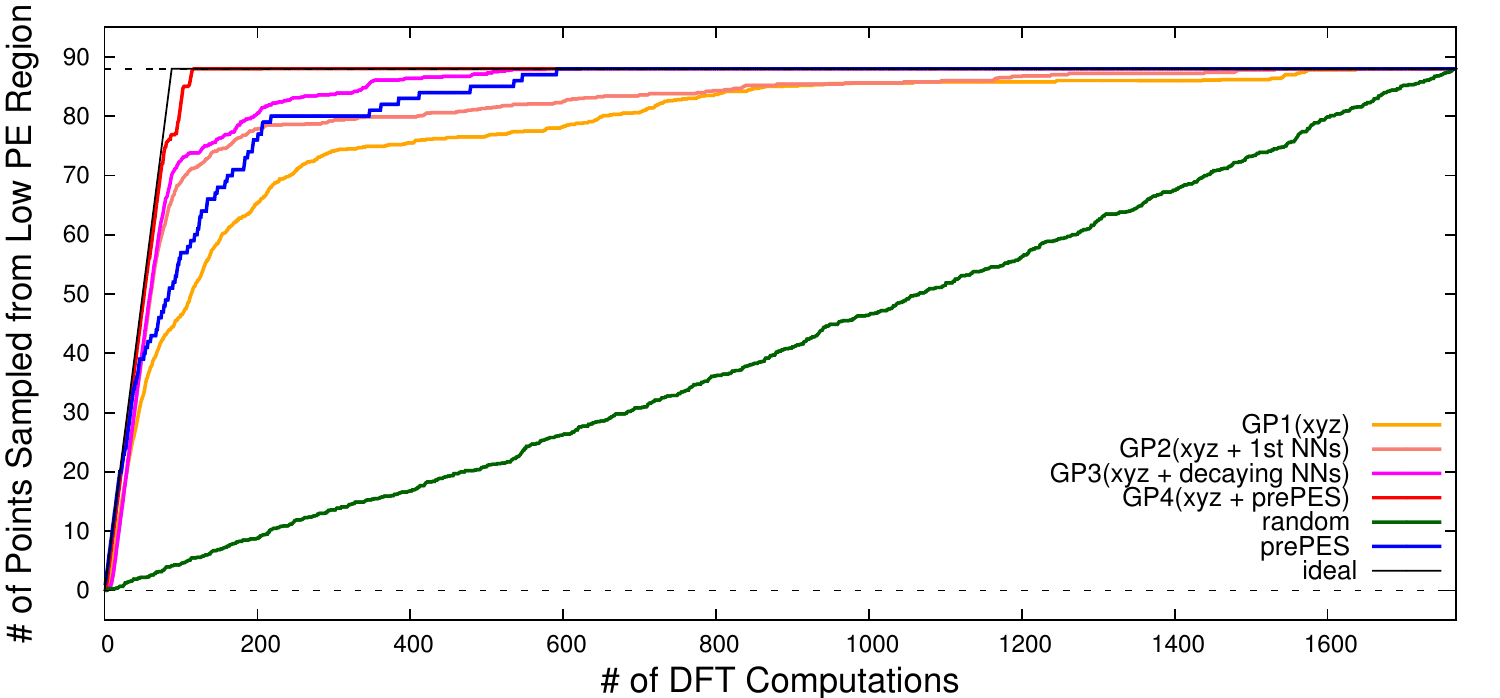} \\
(a) $\alpha=0.05$ \\
\includegraphics[width=0.75\textwidth]{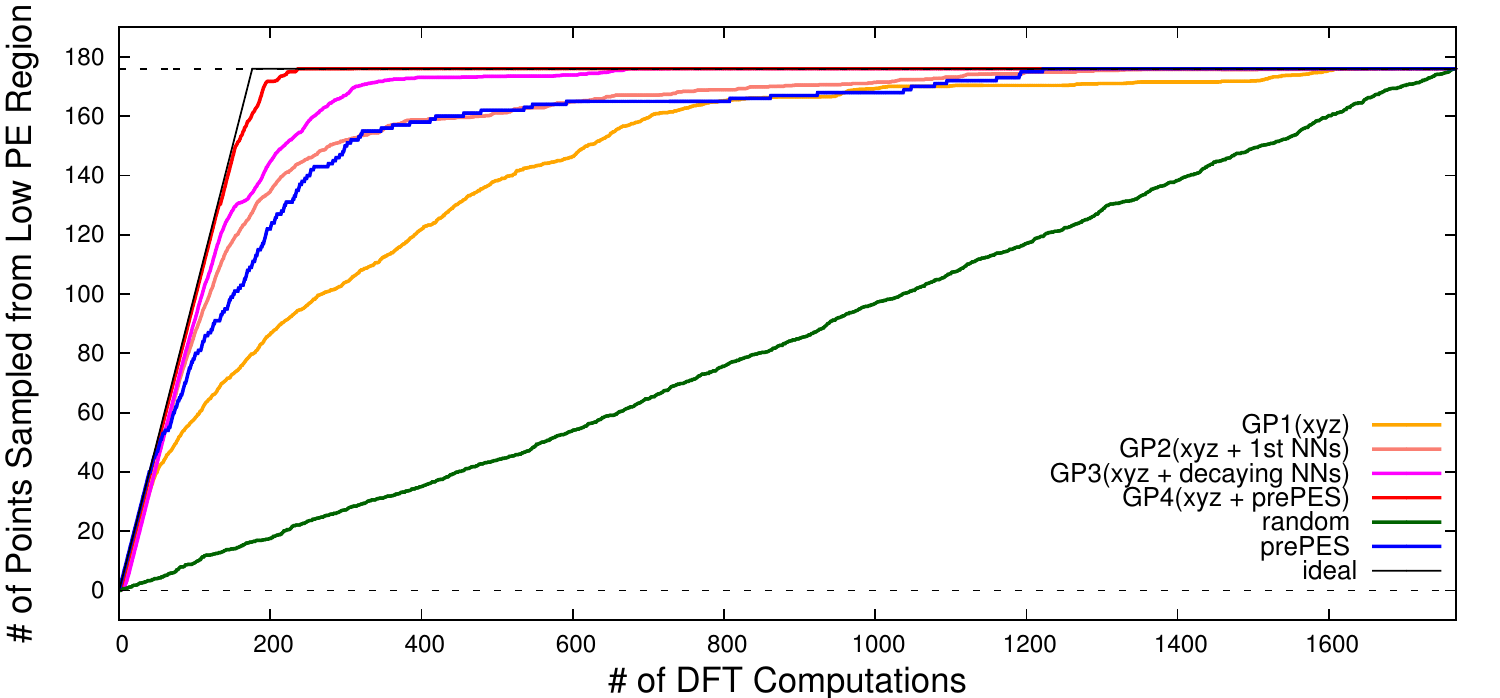} \\
(b) $\alpha=0.10$ \\
\includegraphics[width=0.75\textwidth]{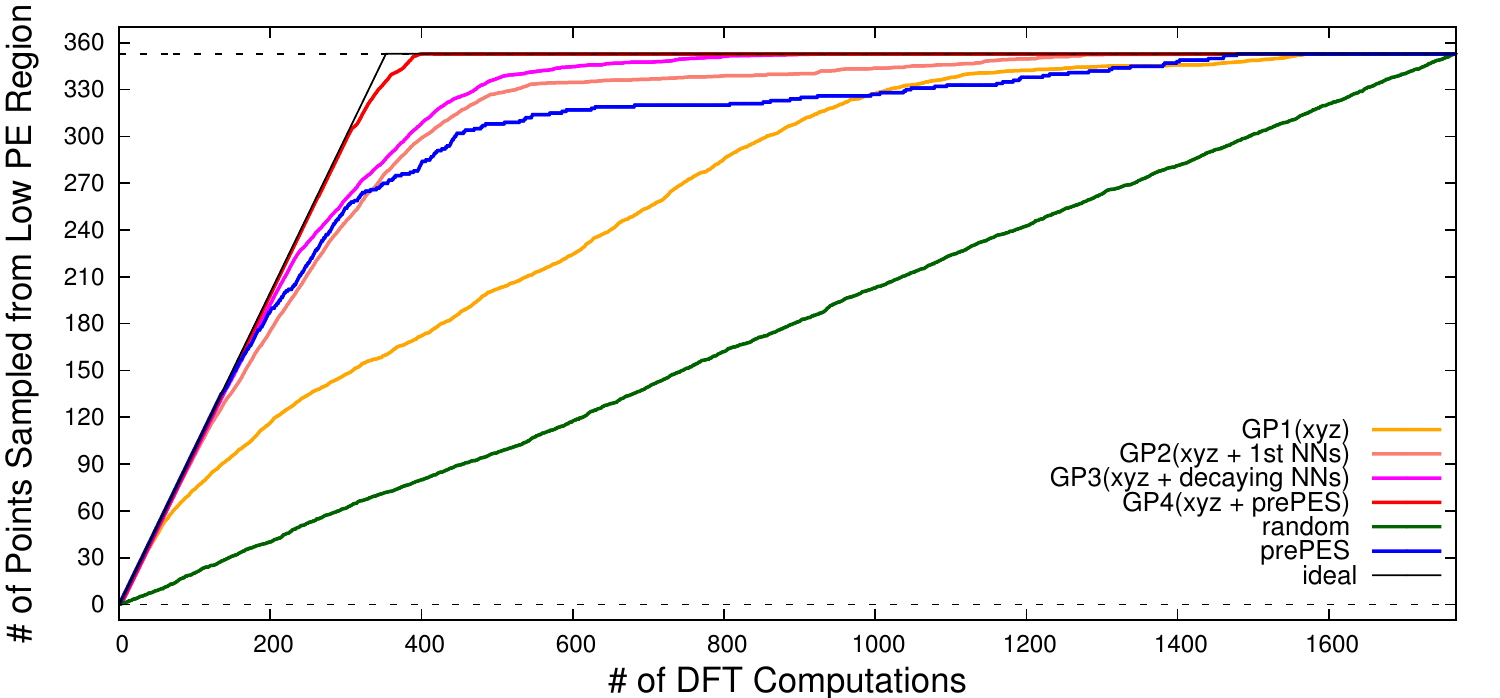} \\
(c) $\alpha=0.20$ \\
\caption{
 Efficiencies of the seven sampling methods for
 (a) $\alpha = 0.05$,
 (b) $\alpha = 0.10$ and
 (c) $\alpha = 0.20$. 
 The numbers of grid points
 successfully sampled from the low PE region 
 (\#(TP))
 are plotted
 as functions of 
 the number of PE evaluations by DFT computations 
 (\#(TP) + \#(FP)). 
}
\label{fig:result-efficiency}
\end{center}
\end{figure*}

\figurename \ref{fig:result-efficiency}
and
Table \ref{tab:result-efficiency}
compare the efficiencies of the seven sampling methods.
In each of the three plots (corresponding to the $\alpha$ = 0.05, 0.1, and 0.2 cases), 
the number of points successfully sampled from the low PE region
(\#(TP))
is plotted
as a function of 
the number of PE evaluations by DFT calculations
(\#(TP) + \#(FP)). 

\begin{table}[t]
 \caption{
 Numbers of sampling steps
 required to identify all the grid points in the low PE region. 
 The averages and the standard deviations
 over 10 runs
 with different random seeds
 are shown.
 Note that
 the results of {\tt prePES} are obtained
 by a single run
 because there is no random component in {\tt prePES}.}
 \label{tab:result-efficiency}
 \begin{center}
  \begin{tabular}{c||c|c|c}
   & $\alpha=0.05$ & $\alpha=0.10$ & $\alpha=0.20$ \\ \hline
   {\tt GP1}& 
   $1488.5 \pm 239.4$ & 
   $1576.1 \pm 36.9$ & 
   $1539.6 \pm 31.2$ 
   \\
   {\tt GP2}& 
   $1294.1 \pm 180.9$ & 
   $1266.9 \pm 167.9$ & 
   $1269.4 \pm 100.3$ 
   \\
   {\tt GP3}& 
   $511.7 \pm 30.4$ & 
   $617.6 \pm 49.7$ & 
   $875.2 \pm 60.1$ 
   \\
   {\tt GP4} & 
   $114.3 \pm 1.3$ &
   $235.5 \pm 1.0$ &
   $394.1 \pm 5.2$
   \\
   {\tt random} & 
   $1749.6 \pm 15.0$ & 
   $1757.1 \pm 11.3$ & 
   $1762.2 \pm 3.9$ 
   \\
   {\tt prePES} & 
   $592$ & 
   $1222$ & 
   $1479$ 
   \\ \hline
   {\tt ideal}  &
   $89$ &
   $177$ &
   $354$ \\	       
  \end{tabular}
 \end{center}
\end{table}

The results for the four different GP-based methods
({\tt GP1} to {\tt GP4})
indicate the importance of the choice of descriptors.
When only the three-dimensional coordinate information 
({\tt xyz})
is used, 
the performances were only slightly better than {\tt random}.
For example,
when $\alpha = 0.20$, 
{\tt GP1} required 1539.6 DFT calculations in average 
until all the points in the low PE region
were successfully identified.
The results on
{\tt GP2} and {\tt GP3}
were better than
{\tt GP1},
suggesting that
adding appropriate descriptors are generally advantageous. 
The results on 
{\tt GP4}
indicate that
a preliminary PES is highly helpful as a descriptor. 
However, 
the results on
{\tt prePES}
suggest that
the preliminary PES alone is not sufficient for efficiently identifying the low PE region. 
In \S\ref{subsec:importance-prePES},
we discuss the importance of the preliminary PES in more detail. 

\begin{figure*}[htbp]
 \begin{center}
  \begin{tabular}{cccc}
   100 steps & 200 steps & 300 steps & 400 steps \\
   \hline
\multicolumn{4}{c}{{\tt GP1} sampling} \\ \hline
\includegraphics[width=.175\textwidth]{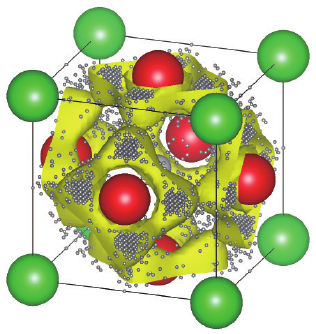} &
\includegraphics[width=.175\textwidth]{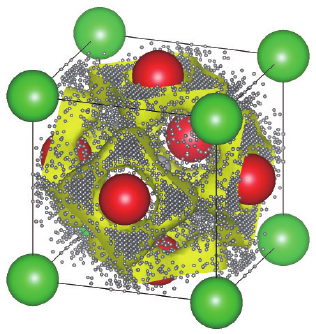} &
\includegraphics[width=.175\textwidth]{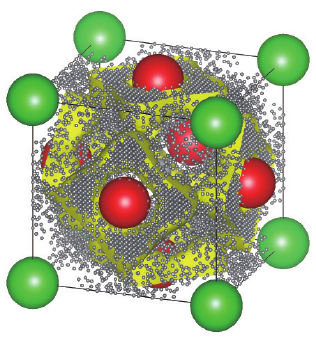} &
\includegraphics[width=.175\textwidth]{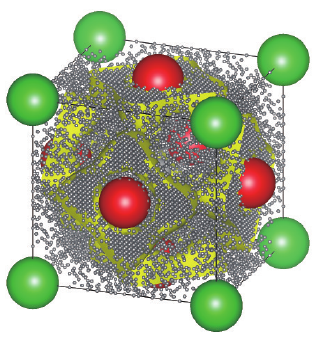} \\
\hline
\multicolumn{4}{c}{{\tt GP4}  sampling} \\ \hline
\includegraphics[width=.175\textwidth]{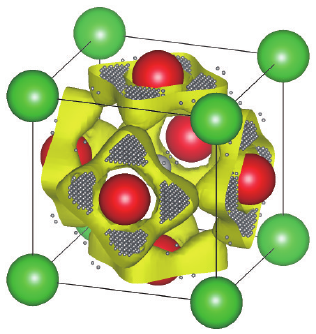} &
\includegraphics[width=.175\textwidth]{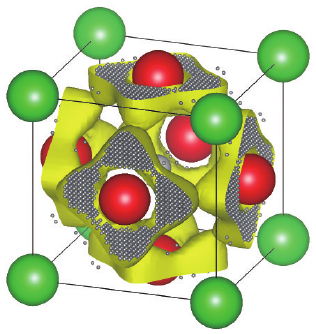} &
\includegraphics[width=.175\textwidth]{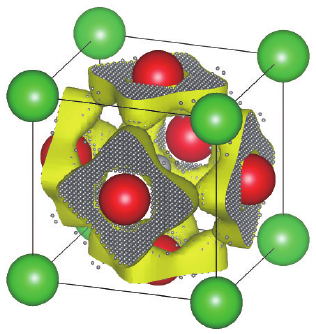} &
\includegraphics[width=.175\textwidth]{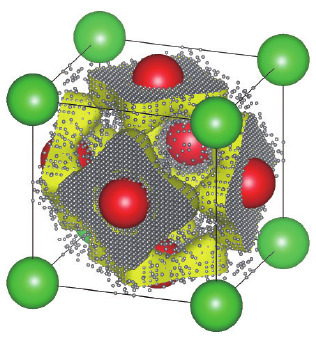} \\
\hline
\multicolumn{4}{c}{{\tt prePES} sampling} \\ \hline
\includegraphics[width=.175\textwidth]{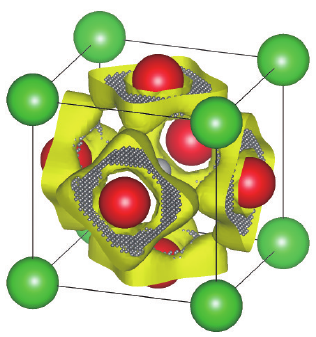} &
\includegraphics[width=.175\textwidth]{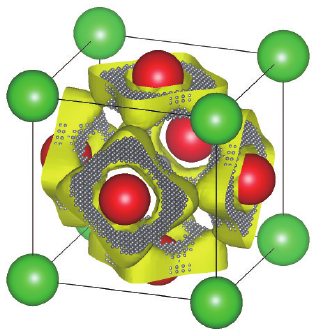} &
\includegraphics[width=.175\textwidth]{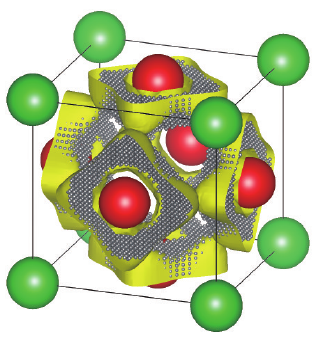} &
\includegraphics[width=.175\textwidth]{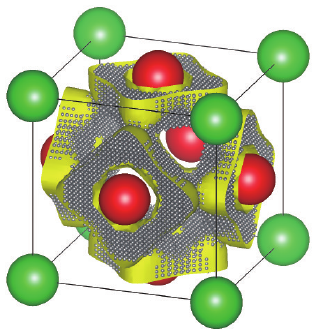} \\
\hline
\multicolumn{4}{c}{{\tt random} sampling} \\ \hline
\includegraphics[width=.175\textwidth]{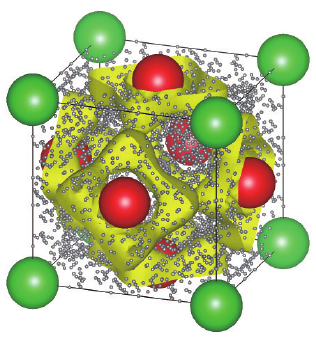} &
\includegraphics[width=.175\textwidth]{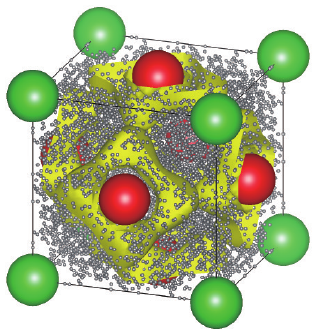} &
\includegraphics[width=.175\textwidth]{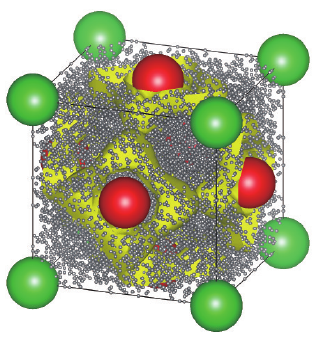} &
\includegraphics[width=.175\textwidth]{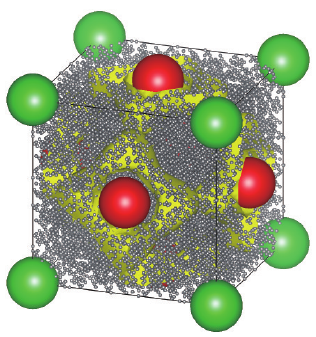} \\
\hline
\multicolumn{4}{c}{{\tt ideal} sampling} \\ \hline
\includegraphics[width=.175\textwidth]{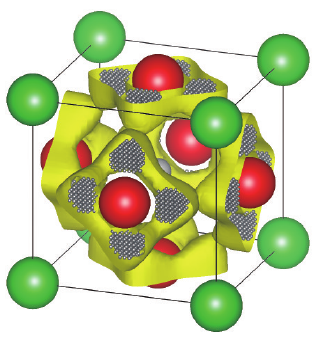} &
\includegraphics[width=.175\textwidth]{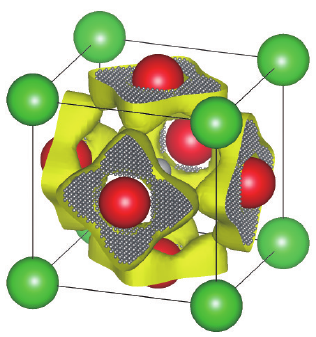} &
\includegraphics[width=.175\textwidth]{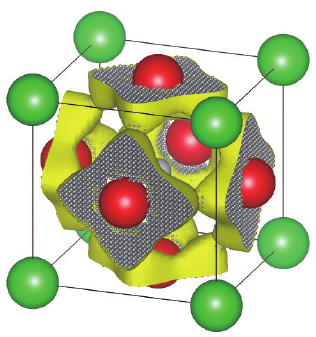} &
\includegraphics[width=.175\textwidth]{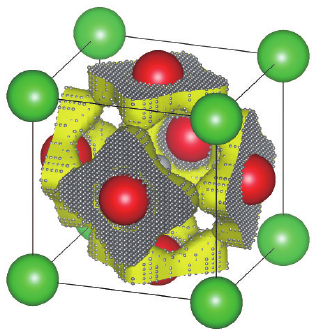} \\
\end{tabular}
\caption{
  Selected grid points
  (gray dots)
  at 100, 200, 300, and 400 steps
  by different sampling methods
  in the model crystal lattice of ${\rm BaZrO_3}$
  for $\alpha = 0.20$ case.
  The yellow surface
  in each plot is the isosurface
  corresponding to the energy threshold at $\alpha = 0.2$.
  }
  \label{fig:sampling-points}
 \end{center}
\end{figure*}

\figurename \ref{fig:sampling-points}
demonstrates the differences between the sampling sequences 
of the 
{\tt GP1}, {\tt GP4}, {\tt prePES}, {\tt random} and {\tt ideal} methods
(we omit those of {\tt GP2} and {\tt GP3} due to space limitations). 
In {\tt GP1},
many points in the high PE region (non-low PE region) were mistakenly selected.
In {\tt GP4},
the points in the low PE region were preferentially selected,
and only a small number of points were mistakenly selected from the high PE region. 
In {\tt prePES},
the method preferentially selected the points in the low PE region,
but it failed to find all of them. 
Note that
the sampling sequence of {\tt GP4} looks almost identical to that in the {\tt ideal} sampling,
despite the low PE region being unknown beforehand.
This indicates that
the GP model in {\tt GP4} could successfully estimate the PES in the low PE region. 

\subsection{Importance of preliminary PES}
\label{subsec:importance-prePES}

\begin{figure*}[htbp]
\begin{center}
\begin{tabular}{cc}
\includegraphics[width=.35\textwidth]{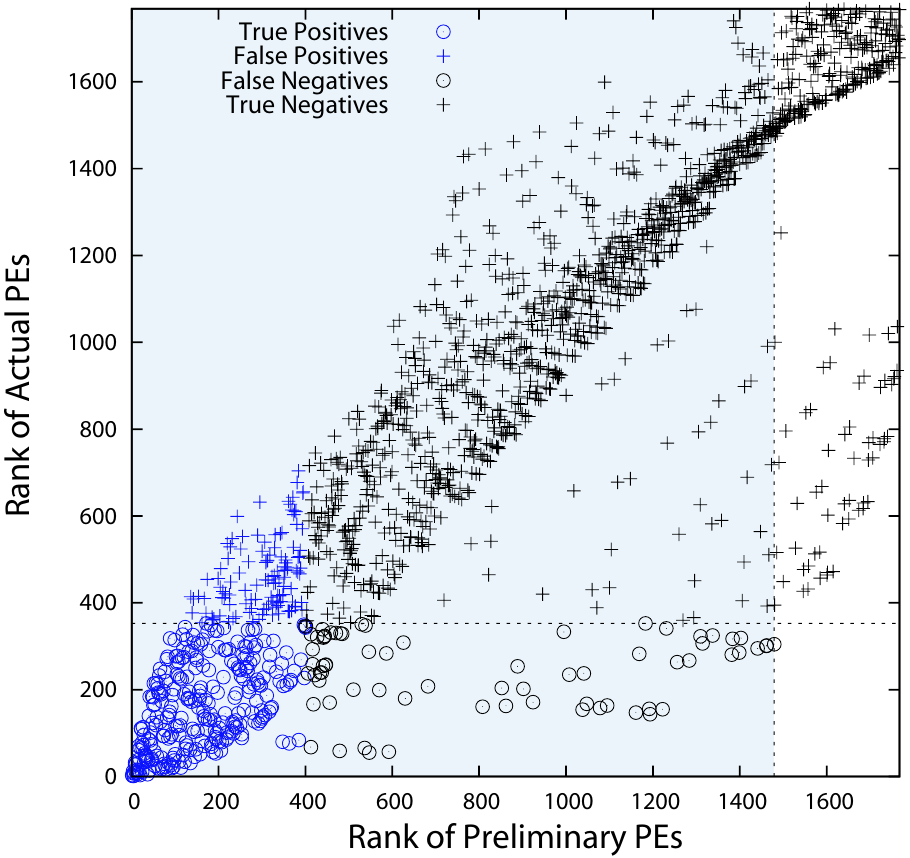} &
\includegraphics[width=.35\textwidth]{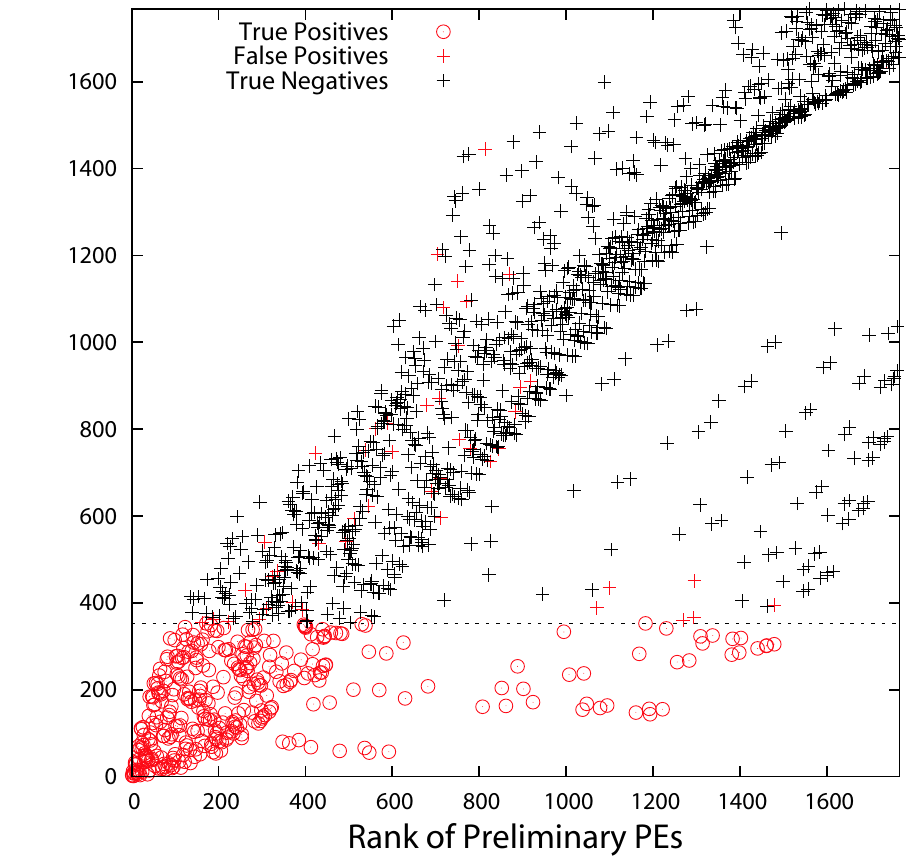} \\
(a) {\tt prePES} &
(b) {\tt GP4} \\
\end{tabular}
\caption{Rank correlation between the actual and the preliminary PEs. 
The grid points in $\cP_\alpha$ are shown as open circles, while those in $\cN_\alpha$ are shown as crosses. 
Sampled points at 400 steps are also shown as blue and red symbols in
 (a) {\tt prePES} and (b) {\tt GP4}, respectively, for the $\alpha = 0.20$ case. 
The method {\tt GP4} samples all the positive points at 400 steps with a small number of false positives, i.e., sampled points that are not included in the low PE region.
Therefore, in (b), there are no false negative points. 
}
\label{fig:rank-correlation}
\end{center}
\end{figure*}

The results in 
\figurename \ref{fig:result-efficiency}
and
Table \ref{tab:result-efficiency}
indicate that
the preliminary PES obtained by single-point DFT calculations
is highly valuable as a descriptor
when it is used along with the three-dimensional coordinates ({\tt xyz})
in GP modeling. 
However, 
using the preliminary PES alone is not sufficient
for identifying the low PE region 
in 
the method 
{\tt prePES}, 
which was only slightly better than
{\tt random}.
Although the sampling curves of
{\tt prePES}
almost overlap with the ideal sampling curves in earlier steps,
they gradually deviate as the sampling proceeds.
Eventually,
the total numbers of sampling steps required
for finding all the points in the low PE region
were
as large as 592, 1222, and 1479 points
for
the $\alpha$ = 0.05, 0.10, and 0.20 cases respectively,
which are 6.7-fold, 6.9-fold, and 4.2-fold worse than the {\tt ideal} sampling cases. 
This inefficiency of {\tt prePES} is ascribed to the relationship
between the DFT calculation with structural optimization and the single-point DFT calculation.
This can be clearly seen in the rank correlation
between the actual and preliminary PEs
shown in \figurename \ref{fig:rank-correlation}.
\figurename \ref{fig:rank-correlation}
shows
the rank correlation for
the $\alpha = 0.20$ case,
where 
points with low PEs are located below the horizontal broken line.
Since the
{\tt prePES}
sampling method selects the points in the increasing order of the preliminary PEs,
meaning that the points are selected from left to right in \figurename \ref{fig:rank-correlation},
most of the $N$ grid points (all points located in the shaded region)
must be sampled for selecting all the points in the low PE region.

\begin{figure*}[htbp]
 \begin{center}
  \begin{tabular}{cc}
   \includegraphics[width=.4\textwidth]{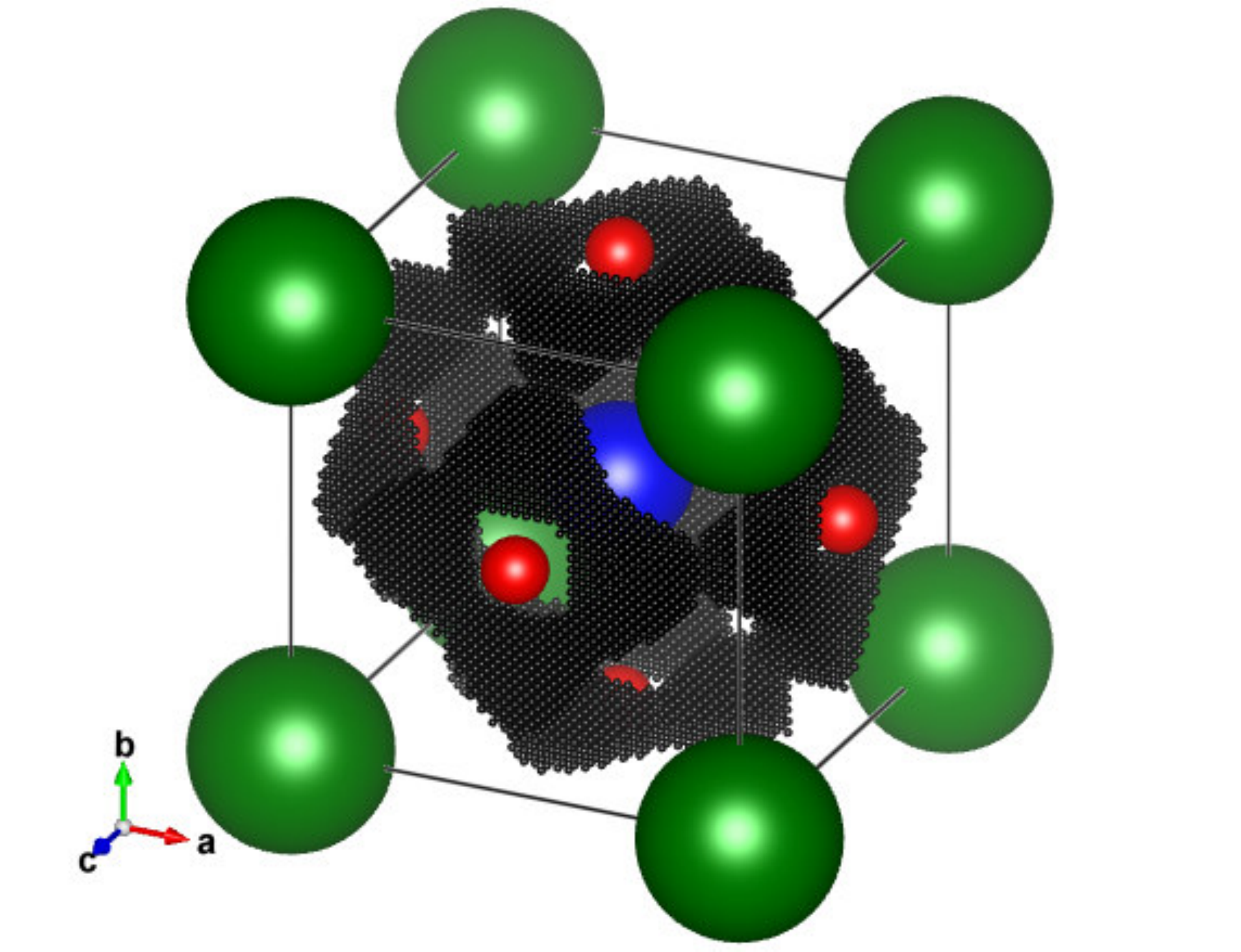} &
   \includegraphics[width=.4\textwidth]{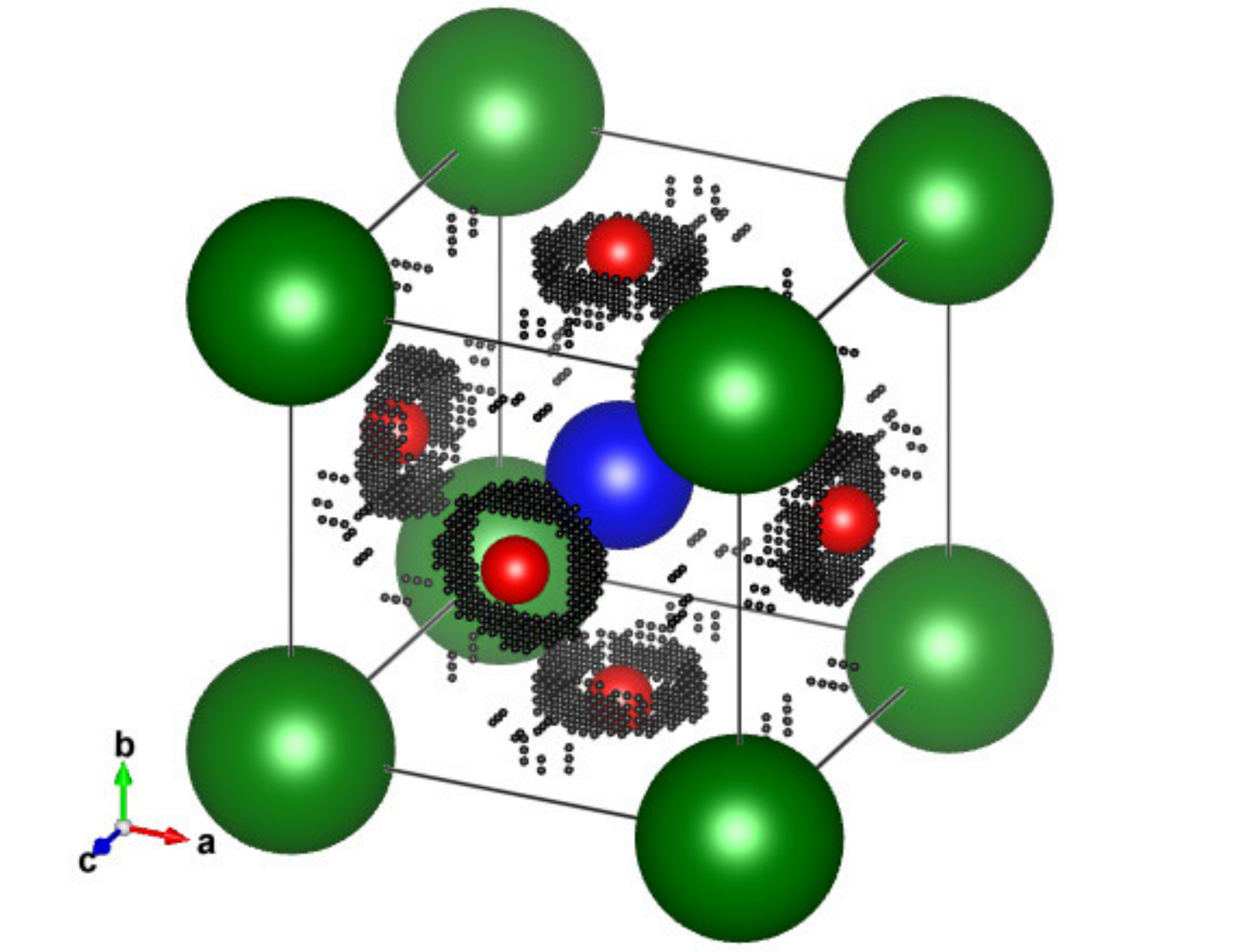} \\
   (a) Grid points in the low PE region with $\alpha = 0.2$ &
   (b) Grid points that could not be found by {\tt prePES}
  \end{tabular}
  \caption{
  (a) Grid points
  in the low PE region
  with
  $\alpha = 0.2$
  of
  ${\rm BaZrO_3}$, 
  and
  (b)
  grid points
  that could not be found
  via the {\tt prePES} sampling method 
  in 394 steps
  by which
  the {\tt GP4} sampling method could successfully identify all the grid points in the low PE region. 
  Plot (b) indicates that 
  grid points around the O atom (red) could not be correctly identified
  by using preliminary PES alone,
  i.e., 
  the preliminary PEs
  at these points are NOT relatively low, 
  although
  their actual PEs are found to be relatively low 
  after structural optimization. 
  }
  \label{fig:points-not-found}
 \end{center}
\end{figure*}

\figurename \ref{fig:points-not-found}
shows
the points
that could not be found by
the {\tt prePES} sampling method. 
This reflects the displacement of the first-nearest-neighbor O ion (${\rm O_{\rm 1NN}}$) for keeping a suitable distance from the proton ($\sim$ 1 \AA) in the case of ordinary DFT calculation with structural relaxation.
In addition, the structural relaxation around a proton critically affects the potential barrier heights for the proton migration.
The potential barrier of the rotational and hopping paths obtained by the single-point DFT calculations are 0.37 and 1.29 eV, respectively, which are much higher than the ground truth values obtained by the ordinary fully relaxed DFT calculations.

%
%

On the other hand,
by using GP with {\tt xyz} and {\tt prePES} as descriptors, 
the points close to O atom
were also successfully identified. 
This is due to the advantage of the GP-based procedure 
where
we could successfully predict
by using the GP model
that
the actual PE values
and
the preliminary PE values
are highly different in the neighborhoods of O atoms. 
In
{\tt GP4}, 
the average number of sampling steps 
required for identifying all the points in the low PE region 
was only 394.1 
which is only 1.1-fold of the {\tt ideal} sampling method. 
\section{Practical Issues}
\label{sec:practical-issue}

In this section,
we present results regarding two practical issues.
In \S \ref{subsec:stopping-criterion}, 
we show results on when the sampling should be stopped.
In \S \ref{subsec:select-alpha}, 
we discuss and present results on how an appropriate $\alpha$ can be selected.

\subsection{Stopping criterion}
\label{subsec:stopping-criterion}
\begin{figure}[htbp]
 \begin{center}
   \includegraphics[width=.50\textwidth]{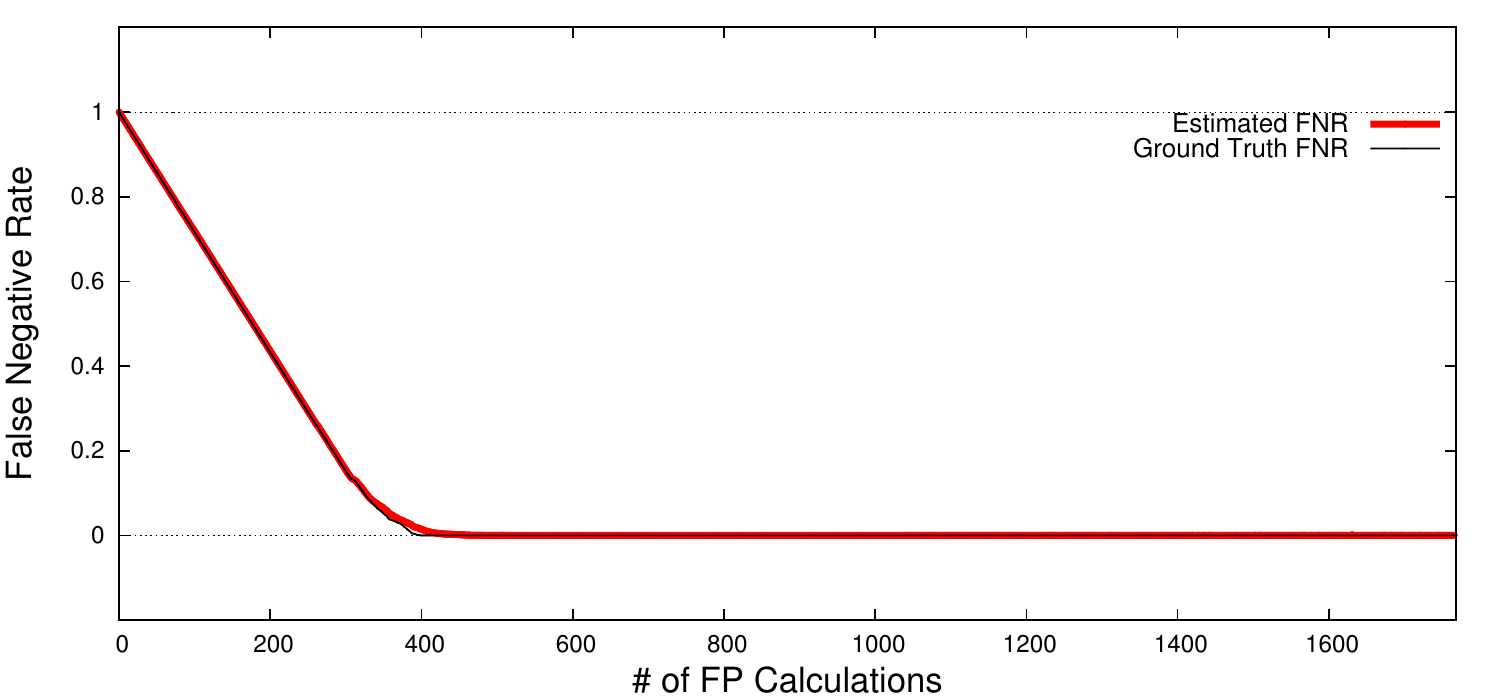} \\
   (a) The estimated FNR \\
   \includegraphics[width=.50\textwidth]{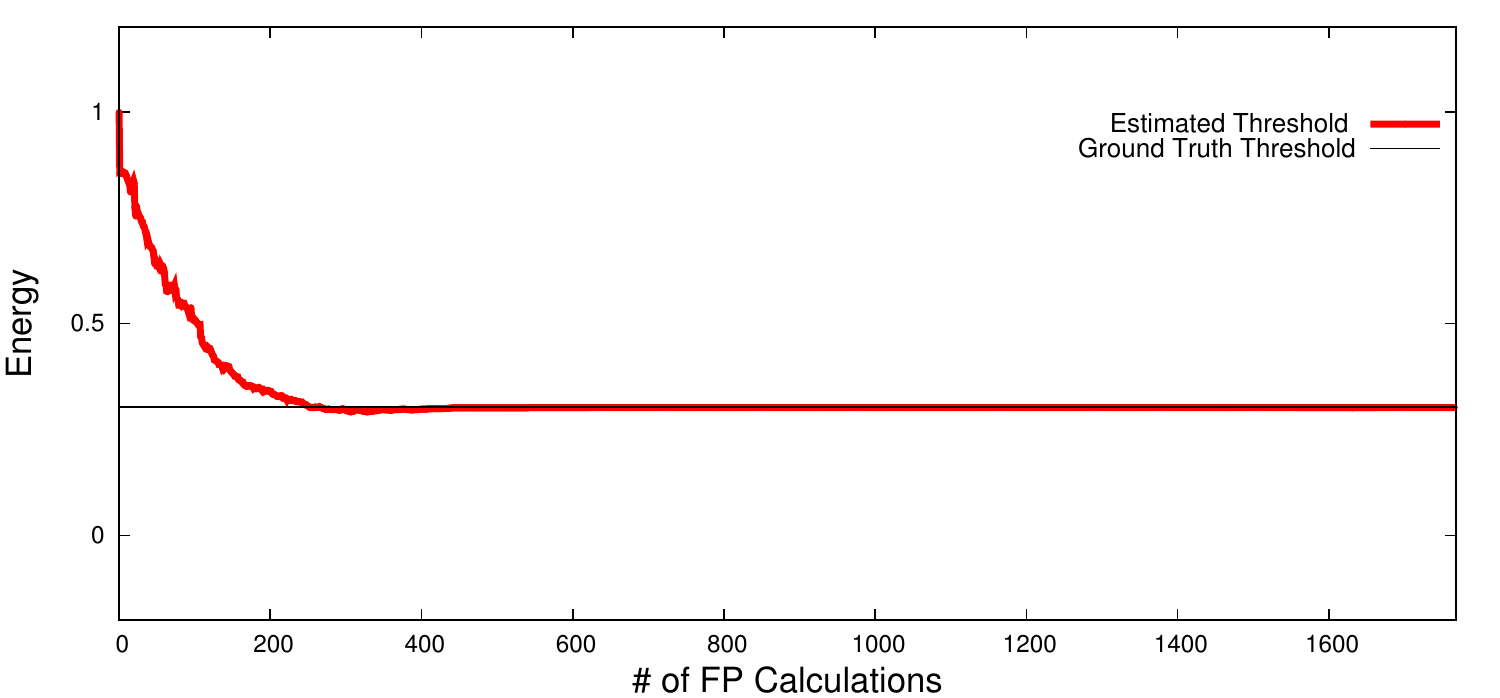} \\
   (b) The estimated threshold \\
   \caption{The profiles of the estimated FNRs and thresholds for $\alpha = 0.20$ case.}
   \label{fig:stopping-condition}
 \end{center}
\end{figure}

One of the important practical issues in selective sampling methods is when the sampling should be stopped.
As we discussed in
\S \ref{subsec:key-feature3}, 
one of the practical advantages of statistical models such as the GP model is that we can estimate how many points still remain to be sampled by estimating the false negative rate (FNR). \figurename \ref{fig:stopping-condition} shows the profiles of the estimated FNR and the estimated threshold as functions of the number of DFT calculations in the case of $\alpha$ = 0.20. These plots indicate that the estimated FNR almost coincides with the ground truth line, and that the estimated threshold also converges to the true value as the sampling proceeds. These results suggest that the estimated FNR would be actually useful as a stopping criterion.

\subsection{How to determine an appropriate $\alpha$ value}
\label{subsec:select-alpha}

Another practical issue is how to choose an appropriate $\alpha$ depending on what physical property is being investigated. In the case of proton conduction in an oxides, we should identify the low PE region in such a way that a proton-conducting network exists throughout the crystal lattice within the region. According to the actual PEs in \figurename \ref{fig:PE-relaxed}, the low PE regions are isolated when $\alpha < 0.15$, while they are abruptly connected when $\alpha \simeq 0.20$. This means that a proper $\alpha$ value would be around 0.20 in the present study.

\begin{figure}[htbp]
 \begin{center}
   \includegraphics[width=.50\textwidth]{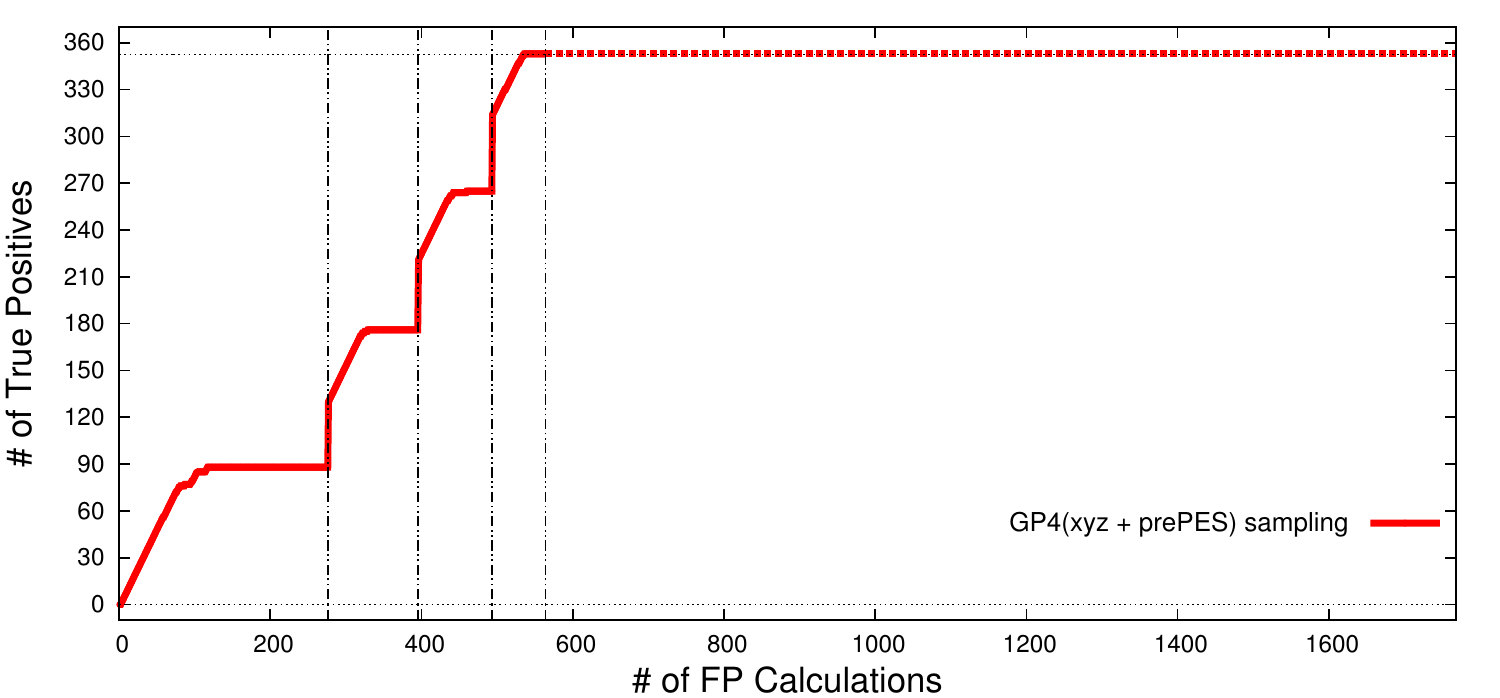} \\
   (a) Sampling profile  \\
   \includegraphics[width=.50\textwidth]{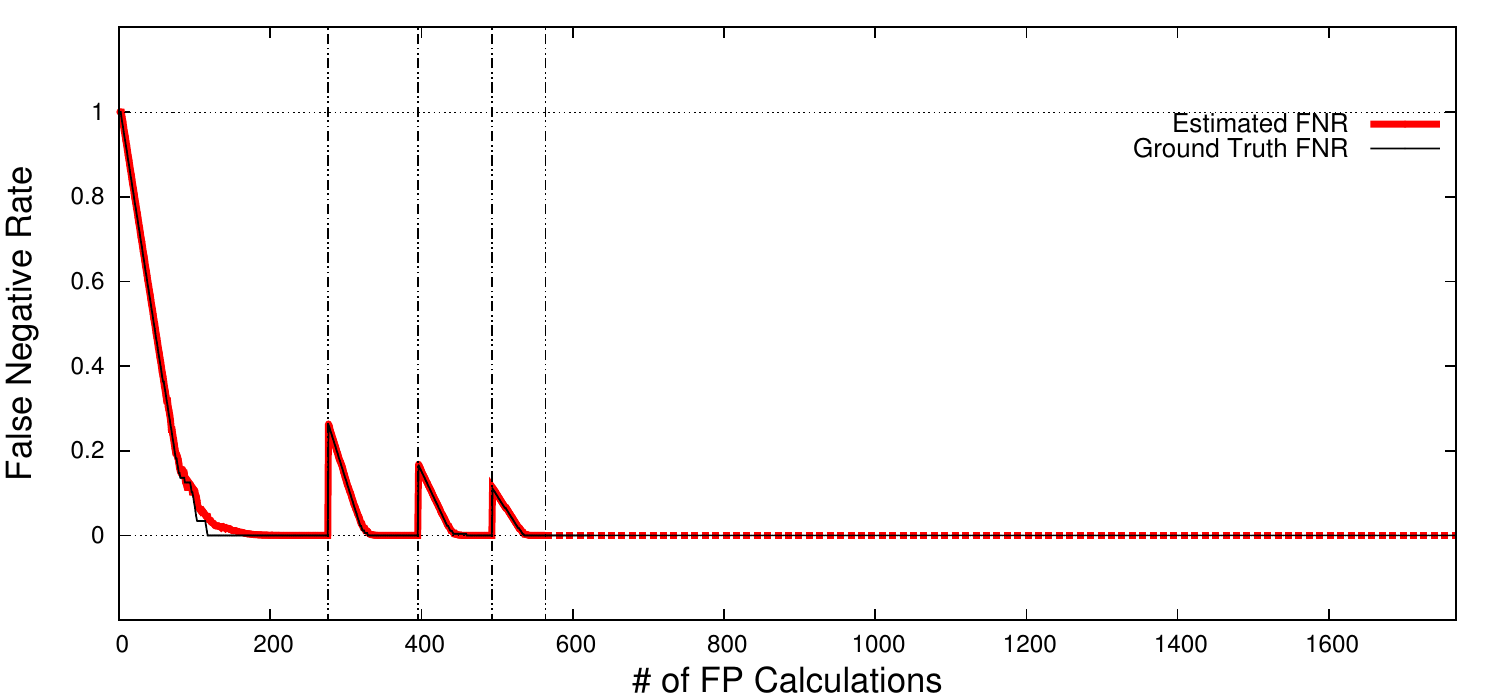} \\
   (b) FNR \\
   \includegraphics[width=.50\textwidth]{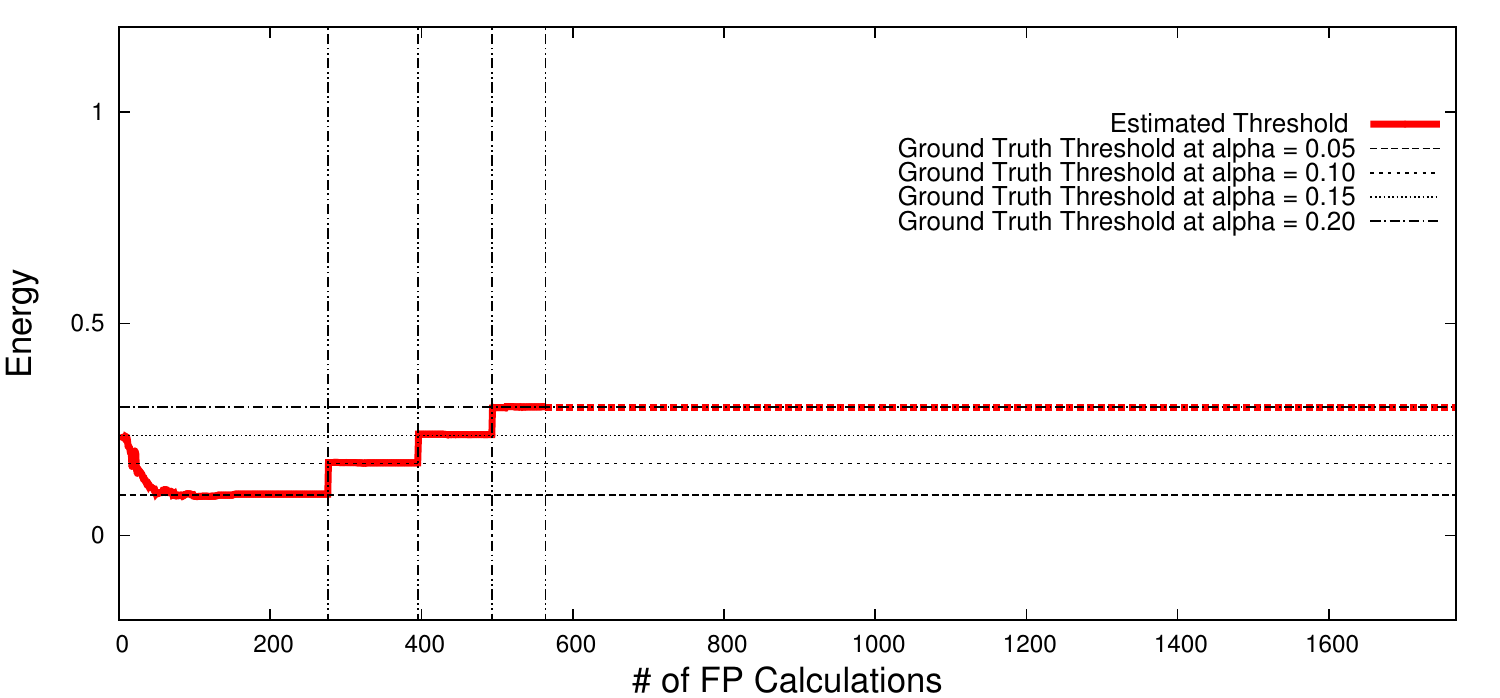} \\
   (c) Threshold \\
   \caption{Results when $\alpha$ is gradually increased as 0.05 $\to$ 0.10 $\to$ 0.15 $\to$ 0.20 in a stepwise manner. (a) THe true positive points, (b) the estimated FNR, and (c) the estimated threshold are plotted as functions of the numbers of DFT computations. Here, $\alpha$ is updated when the estimated false negative rate $\hat{\rm FNR}$ becomes less than $\veps = 10^{-6}$ at the current $\alpha$. Note that the FNR in (b) and the threshold in (c) change with $\alpha$.}
   \label{fig:select-alpha} \end{center}
\end{figure}

If such an appropriate $\alpha$ value is unknown beforehand, we can set the $\alpha$ value in a stepwise manner. For demonstrating this approach, we investigated the performance of {\tt GP4} when $\alpha$ was increased as 0.05, 0.10, 0.15 and 0.20 in a stepwise manner, whose results are shown in \figurename \ref{fig:select-alpha}. In this scenario, we increased $\alpha$ by 0.05 when the estimated FNR became smaller than $10^{-6}$. The plot (b) indicates that, in the first step with $\alpha = 0.05$, the convergence of the estimated FNR was a bit slower than the ground truth FNR. This is why we had to sample more than 250 times before we were sure that all the points in $\cP_{0.05}$ had been successfully sampled. On the other hand, when $\alpha$ = 0.10, 0.15, and 0.20, the convergences of the FNRs were almost as fast as the ground truth FNRs. Note that the true positive points abruptly increase when the $\alpha$ value is switched, indicating that the positive points for higher $\alpha$ have already been sampled in earlier steps. Although this stepwise strategy was a little less efficient than directly specifying $\alpha = 0.20$, it was still much more efficient than the {\tt prePES} and {\tt random} sampling methods.

\section{Conclusion}
\label{sec:conclusion}


We introduced a machine learning-based selective sampling procedure for PES evaluation,
and demonstrated the efficacy of the proposed procedure
by applying it to proton conduction in
${\rm BaZrO_3}$.
The performance of the selective sampling method
based on the GP model greatly
depends on the descriptors,
and the use of a preliminary PES
({\tt GP4(xyz + prePES)}) is significantly effective,
which was here evaluated by single-point DFT calculations in a smaller supercell.
This indicates that
the machine learning approach hybridized with a low-cost PES evaluation is a solid methodology for preferential PES evaluation in a region of interest.
In addition,
we demonstrated that two practical issues,
namely, when to stop the sampling and how to determine an appropriate $\alpha$ value (equivalent to the PE threshold),
can be solved by using the false negative rate (FNR) defined in Eq. (12).

\begin{acknowledgements}
 KT was partially supported by JSPS KAKENHI 25106002.  AS was partially supported by JSPS KAKENHI 15H04116 and 25106005.  MS was partially supported by JSPS KAKENHI 26106510. AK was partially supported by JSPS KAKENHI 25106008. IT was partially supported by JSPS KAKENHI 26280083 and 26106513, and by JST CREST 13418089. 
\end{acknowledgements}

%


\end{document}